\documentclass{aa}  
\pdfoutput=1
\usepackage{natbib}
\usepackage{graphicx}
\usepackage[normalem]{ulem}

\usepackage{float}

\usepackage{txfonts}
 \usepackage{hyperref}
\newcommand{\kms}{\,km$\,$s$^{-1}$}	

\begin{document}

   \title{Binarity among CEMP-no stars: \\ an indication of multiple formation pathways? \thanks{based on observations made with the Southern African Large Telescope (SALT) and the Canada-France-Hawaii Telescope (CFHT)}}

   \author{A. Arentsen
          \inst{1}
          \and
          E. Starkenburg\inst{1}
          \and
          M. D. Shetrone\inst{2}
          \and
          K. A. Venn\inst{3}
          \and
          \'E. Depagne\inst{4}
           \and
          A. W. McConnachie\inst{5}
          }

   \institute{Leibniz-Institut f\"ur Astrophysik Potsdam (AIP), An der Sternwarte 16, D-14482 Potsdam, Germany
         \and
             McDonald Observatory, University of Texas at Austin, HC75 Box 1337-MCD, Fort Davis, TX 79734, USA
         \and 
            Department of Physics \& Astronomy, University of Victoria, Victoria, BC, V8W 3P2, Canada
         \and 
            South African Astronomical Observatory (SAAO), Observatory Road Observatory Cape Town, WC 7925, South Africa
          \and
             NRC Herzberg Institute of Astrophysics, 5071 West Saanich Road, Victoria, BC V9E 2E7, Canada
         }

   \date{Received XXX; accepted YYY}

 
  \abstract{Carbon-enhanced metal-poor (CEMP) stars comprise a large percentage of stars at the lowest metallicities. The stars in the CEMP-no subcategory do not show any $s$-process enhancement and therefore cannot easily be explained by transfer of carbon and $s$-process elements from a binary AGB companion. We have performed radial velocity monitoring of a sample of 22 CEMP-no stars to further study the role binarity plays in this type of CEMP star. We find four new binary CEMP-no stars based on their radial velocity variations, thereby significantly enlarging the population of known binaries to a total of eleven. One of the new binary systems is HE~0107$-$5240, one of the most iron-poor stars known, supporting the binary transfer model for the origin of the abundance pattern of this star.
In our sample we find a difference in binary fraction depending on the absolute carbon abundance, with a binary fraction of $47^{\,+15\,}_{\,-14}\%$ for stars with higher absolute carbon abundance and $18^{\,+14\,}_{\,\,-9} \%$ for stars with lower absolute carbon abundance. This potentially implies a relation between a high carbon abundance and the binarity of a metal-poor star. Although binarity does not equate to mass transfer, there is a possibility that a CEMP-no star in a binary system has been polluted and care has to be taken in the interpretation of their abundance patterns. We furthermore demonstrate the potential of Gaia to discover additional binary candidates. }


   \maketitle


\section{Introduction}

To study the earliest times in the Universe we do not have to go to high redshift. Our Milky Way still hosts remnants from these early times in the form of extremely metal-poor stars that are expected to be almost as old as the Universe and which we can study in detail. At the lowest metallicities, the fraction of stars enhanced in carbon increases dramatically (\citealt{Beers92}; \citealt{Norris97}). These carbon-enhanced metal-poor (CEMP) stars comprise $15-20\%$ of the very metal-poor stars ([Fe/H]\footnote{[X/Y] $ = \log(N_\mathrm{X}/N_\mathrm{Y})_* - \log(N_\mathrm{X}/N_\mathrm{Y})_{\odot}$, where the subscript * refers to the considered star, and N is the number density.} $<-2.0$), which increases to $\sim40\%$ for extremely metal-poor stars ([Fe/H] $<-3.0$) and even higher percentages at lower metallicities (\citealt{Yong13}; \citealt{Lee13}; \citealt{Placco14}). 

There are different types of CEMP stars initially defined by \citet{BeersChristlieb05}, the two main classes being CEMP-$s$ stars that show additional enhancement in s-process elements (with [C/Fe] $> +0.7$ and [Ba/Fe]~$> +1.0$), and the CEMP-no stars that do not show any s-process enhancement and which usually occur at lower metallicities (with [C/Fe] $> +0.7$ and [Ba/Fe] $ < 0.0$). A subclass of the CEMP-$s$ stars are the CEMP-$r/s$ stars that are additionally enhanced in r-process elements. It was noticed by \citet{Spite13} that CEMP stars seemed to occupy two bands in absolute carbon versus metallicity space. The more metal-rich CEMP stars have higher absolute carbon clustering around an absolute carbon abundance A(C)\footnote{A(X) $= \log \epsilon_\mathrm{X} = \log(N_\mathrm{X}/N_\mathrm{H}) + 12$. Throughout this paper we assume \citet{Asplund09} solar abundances.} $\sim 8.25$ and they turned out to be mainly CEMP-$s$ stars, whereas the more metal-poor CEMP stars are located at a lower A(C)~$\sim 6.5$ and they are mainly CEMP-no stars. Larger samples of CEMP stars have confirmed this trend (e.g. \citealt{Bonifacio15}; \citealt{Hansen15}), although there are always some outliers.

Through radial velocity monitoring it was found that the CEMP-$s$ stars are almost always in a binary system (e.g. \citealt{McClureWoodsworth90}; \citealt{PrestonSneden01}; \citealt{Lucatello05}; \citealt{Hansen16b}), while the CEMP-no stars more often appear to be single stars (\citealt{Norris13b}; \citealt{Starkenburg14}; \citealt{Hansen16a}, afterwards S14 and H16a). CEMP-$s$ stars are thought to have received their carbon and s-process elements via mass-transfer from an evolved companion that has gone through the Asymptotic Giant Branch (AGB) phase \citep{Abate15}. 

The exact origin of the CEMP-no stars is not yet clear. CEMP-no stars are not generally considered to be in binary systems, but the data indicate that at least some of them are: $\sim 17\%$ of the sample in H16a. This is close to the binary frequency of $16\% \pm 4\%$ found by \citet{Carney03} for 91 carbon-normal metal-poor ([Fe/H] $\leq -1.4$) field red giants. Since most of the CEMP-no stars do not have a binary companion, it is often assumed that the carbon abundance in these stars is intrinsic and therefore reflects the composition of the gas out of which they are formed. The thirteen stars with the lowest known metallicities ([Fe/H] $< -4.5$) are all CEMP stars with two exceptions from \citet{Caffau11} and (likely) \citet{Starkenburg18}, additionally most of them do not show clear signatures of s-process enhancement. This combination of the most metal-poor stars being enhanced in carbon and not in s-process elements suggests their abundances may be ``original". The CEMP-no stars may be early-generation stars born from gas polluted by the first generation(s) of massive stars. 

One of the possible progenitors of carbon in the early universe are the so-called spinstars (e.g. \citealt{Meynet06} \citeyear{Meynet10}; \citealt{Chiappini13}). They are rapidly rotating massive ultra metal-poor stars with strong winds, and they can form large amounts of carbon. Another possibility for the progenitors of carbon are the so-called faint supernovae with mixing-and-fallback models (\citealt{UmedaNomoto03}; \citeyear{UmedaNomoto05}; \citealt{Nomoto13}; \citealt{Tominaga14}), in which a supernova does not have sufficient energy to eject all its material into its surroundings, but only the outer layers with the lightest elements are ejected while the inner part falls back onto the neutron star or black hole at the centre. Recent work by \citet{Yoon16} suggests that there are two types of CEMP-no stars based on their absolute carbon abundance, possibly corresponding to the two different progenitors.

It is also possible that some CEMP-no stars have been polluted by a companion, but the binary fraction of CEMP-no stars is not yet well constrained. Studying the CEMP-no binary fraction and binary properties of the population provides us with more information on star formation processes at early times. Additionally, knowledge about the binarity of each individual CEMP-no star is important because it may aid the interpretation of the chemical properties of the star. Key in determining the binarity of stars is radial velocity monitoring, a laborious effort. In this paper we present the results of a large radial velocity monitoring program for CEMP-no stars. The initial sample (described in S14) has been extended with additional spectra for 22 CEMP-no stars, including nine new stars that are not in S14 or H16a, which are mainly located in the southern hemisphere. 

This paper is organised as follows. The new observations from this work are described in Section~\ref{sec:data}. In Section~\ref{sec:results} we present the results of the radial velocity monitoring and in Section~\ref{sec:binarysample} we summarise the properties of the CEMP-no binary population. We discuss the results in Section~\ref{sec:discussion}, give an outlook about what can be achieved with \textit{Gaia }in Section~\ref{sec:Gaia} and our conclusions are briefly summarised in Section~\ref{sec:conclusions}.

\section{Data}
\label{sec:data}

\begin{table*}
\caption{\label{table:programstars}The 22 program stars }
\begin{tabular}{lllrllrrrrrl}
 \hline
 Name & \multicolumn{2}{c}{n\tablefootmark{a}} & V & RA & DEC  & [Fe/H] & [C/Fe]\tablefootmark{b} & A(C)\tablefootmark{b} & [Ba/Fe] & ref & bin?\tablefootmark{c}\\
  &  &  & (mag) & (J2000) & (J2000) &  & & & & & \\
 \hline
BD$+44^{\circ}$493 &         C &     5 &   9.1 &  02 26 49.7 &  $+$44 57 47 &  $-3.83$ &    1.35 &  5.95 &    $-0.60$ &        2 &            1 \\    
BS 16929$-$005 &         C &     4 &  13.6 &  13 03 29.5 &  $+$33 51 09  &  $-3.34$ &    0.99 &  6.09 &   $-0.41$ &        1 &            1 \\    
CS 22878$-$027 &         C &     4 &  14.4 &  16 37 35.9 &  $+$10 22 08&  $-2.51$ &    0.86\tablefootmark{d} &  6.78\tablefootmark{d} &  $<-0.75$ &  1 &  1 \\    
CS 22949$-$037 &         C &     5 &  14.4 &  23 26 29.8 &  $-$02 39 58  &  $-4.38$ &    1.16 &  5.97 &    $-0.60$ &        3 &            1 \\    
CS 22957$-$027 &         C &     5 &  13.6 &  23 59 13.1 &  $-$03 53 48 &  $-3.19$ &    2.61 &  7.87 &   $-0.81$ &        3 &           2 \\    
CS 29498$-$043 &         S &     4 &  13.7 &  21 03 52.1 &  $-$29 42 50  &  $-3.87$ &    2.75 &  7.62 &   $-0.49$ &        3 &            1 \\    
CS 29502$-$092 &         C &     6 &  11.9 &  22 22 36.0 &  $-$01 38 28  &   $-3.30$ &    1.06 &  6.59 &   $-1.36$ &        3 &            1 \\    
HE 0057$-$5959 &         S &     2 &  15.8 &  00 59 54.1 &  $-$59 43 30  &  $-4.08$ &    0.86 &  5.21 &   $-0.46$ &        1 &             \\    
HE 0107$-$5240 &         S &     4 &  15.1 &  01 09 29.2 &  $-$52 24 34 &  $-5.44$ &    3.97 &  7.03 &   $< +0.93$ &        4 &  \textbf{3} \\    
HE 0557$-$4840 &         S &     3 &  15.5 &  05 58 39.3 &  $-$48 39 57  &  $-4.73$ &    1.59 &  5.29 &      $< +0.07$ &        5 &               \\    
HE 1012$-$1540 &       C/S &   5/1 &  14.0 &  10 14 53.5 &  $-$15 55 53  &  $-4.17$ &    2.40 &  6.67 &   $-0.28$ &        3 &            1 \\    
HE 1150$-$0428 &         C &     7 &  14.9 &  11 53 06.6 &  $-$04 45 03 &  $-3.47$ &    2.37 &  7.35 &   $-0.48$ &        1 &           2 \\    
HE 1201$-$1512\tablefootmark{e} &       C/S &   4/2 &  13.8 &  12 03 37.1 &  $-$15 29 32  &  $-3.86$ &    1.14 &  5.71 &  $<+0.05$ &  1 &  1 \\    
HE 1300$+$0157 &         C &     4 &  14.1 &  13 02 56.2 &  $+$01 41 52  &  $-3.75$ &    1.31 &  5.99 &  $<-0.85$ &        1 &            1 \\    
HE 1327$-$2326 &         S &     3 &  13.6 &  13 30 05.9 &  $-$23 41 50  &  $-5.71$ &    4.18 &  6.90 &   $< +1.39$ &        6 &            1 \\    
HE 1506$-$0113 &       C/S &   7/1 &  14.4 &  15 09 14.3 &  $-$01 24 57  &  $-3.54$ &    1.47 &  6.38 &    $-0.80$ &        1 &           2 \\    
HE 2139$-$5432 &         S &     2 &  15.4 &  21 42 42.5 &  $-$54 18 43 &  $-4.02$ &    2.59 &  7.01 &  $<-0.33$ &        1 &  \textbf{3} \\    
HE 2142$-$5656 &         S &     3 &  13.7 &  21 46 20.5 &  $-$56 42 18  &  $-2.87$ &    0.95 &  6.61 &   $-0.63$ &        1 &             \\    
HE 2202$-$4831 &         S &     2 &  15.6 &  22 06 06.0 &  $-$48 16 53  &  $-2.78$ &    2.41 &  8.08 &   $-1.28$ &        1 &             \\    
HE 2247$-$7400 &         S &     1 &  13.3 &  22 51 19.6 &  $-$73 44 21  &  $-2.87$ &    0.70 &  6.58 &   $-0.94$ &        1 &             \\ 
SDSS J0140$+$2344\tablefootmark{e,f} &         C &     4 &  15.1 &  01 40 36.2 &  $+$23 44 58  &  $-4.00$ &    1.13 &  5.56 &  $<+0.34$ & 1 &  \textbf{3} \\ 
SDSS J1422$+$0031 &         S &     3 &  16.3 &  14 22 37.4 &  $+$00 31 05 &  $-3.03$ &    1.70 &  7.11 &   $-1.18$ &        7 &  \textbf{3} \\
 \hline
\end{tabular}
\tablefoot{References: (1) \citet{Yong13}, (2) \citet{Ito13}, (3) \citet{Roederer14}, (4) \citet{Christlieb04}, (5) \citet{Norris07}, (6) \citet{Frebel08}, (7) \citet{Aoki13}\\
\tablefoottext{a}{Number of radial velocity measurements added in this program observed with CFHT (C) or SALT (S)} \\
\tablefoottext{b}{Values corrected for evolutionary status, as reported in \citet{Yoon16}} \\
\tablefoottext{c}{1 = single, 2 = binary, from \citet{PrestonSneden01}, S14 and H16a, \textbf{3} = binary from this work} \\
\tablefoottext{d}{This star is not in the \citet{Yoon16} compilation, we report the [C/Fe] from \citet{Yong13} (without evolutionary correction, which is fine since this is not an evolved star) and we computed the A(C) using the \citet{Asplund09} solar carbon abundance} \\
\tablefoottext{e}{These are the dwarf solutions.} \\
\tablefoottext{f}{Sometimes named 53327$-$2044$-$515 (e.g. in \citealt{Yong13}), but here we use its official SDSS name} \\
}

\end{table*}

\subsection{Sample selection and observations}\label{sec:sample}

In this work, we monitor 22 CEMP-no stars for radial velocity variations. The stars were chosen to be extra follow-up for \citet{Starkenburg14} and to extend the sample to the southern hemisphere. Stars were originally selected from \citet{Norris13b} and then the sample was expanded with additional stars from different literature sources. The list of targeted stars with some of their properties can be found in Table~\ref{table:programstars}. All stars in our sample have [Fe/H] $< -2.5$, and more than $80\%$ of the sample has [Fe/H] $<-3.0$. All stars meet the CEMP criterion [C/Fe] $\geq+0.7$, and more than 70\% of the sample meets the stricter CEMP criterion of [C/Fe] $\geq+1.0$. Almost all stars satisfy the classical CEMP-no criterion by having [Ba/Fe] $\leq 0.0$, within the uncertainties. Two notable exceptions are the hyper metal-poor stars HE~0107$-$5240 and HE~1327$-$2326 that have upper limits on [Ba/Fe] larger than $+0.9$. 
However, \citet{Matsuno17} have revised the CEMP-no definition, taking into account the increasing trend in [Ba/Fe] with [Fe/H] among CEMP-s stars. The new definition is different from the classical definition for stars with [C/Fe] $> +2$, where in this region stars are classified as CEMP-no if they have [Ba/C] $< -2$. Both HE~0107$-$5240 and HE~1327$-$2326 satisfy this revised CEMP-no criterion. Additionally, \citet{Norris13b} pointed out that other chemical properties for these stars are consistent with being CEMP-no. 
One other star that does not satisfy even the new CEMP-no criterion is SDSS~0140$+$2344 ([Ba/Fe] $< +0.34$). This star however has an absolute carbon abundance of 5.6, which is much lower than the typical value for CEMP-$s$ stars, therefore it is more likely to be CEMP-no.

Between August 2013 and April 2015 we gathered a total of 98 high-resolution low signal-to-noise spectra of our target stars, plus  spectra of radial velocity standard stars. We made use of the Echelle SpectroPolarimetric Device for the Observation of Stars (ESPaDOnS, \citealt{Donati03}; \citealt{MansetDonati03}) at the Canada-France-Hawaii Telescope (CFHT), with a resolving power of R~$\sim 68\,000$ covering the wavelength range 370--1050~nm. The other spectrograph used was the High Resolution Spectrograph (HRS, \citealt{Bramall10}; \citealt{Bramall12}; \citealt{Crause14}) at the Southern African Large Telescope (SALT, \citealt{Buckley06}), which we used in its Low-Resolution Mode with R~$\sim 16\,000$. Our observations started during commissioning of HRS. The HRS is a fibre-fed dual beam, white pupil echelle spectrograph that yields two spectra: a blue arm that covers from 370--550~nm and a red arm that covers from 550--890~nm. The number of observed spectra per star for each telescope is indicated in Table~\ref{table:programstars}. 

The stars that were observed with CFHT largely overlap with the sample of H16a, which we could not coordinate because our programs were executed around the same time. We typically have a shorter baseline and a more heterogeneous sample of measurements.

\subsection{Data reduction}

The CFHT spectra were reduced using the dedicated software package \textsc{libre-esprit} \citep{Donati97}, which includes an automatic wavelength correction from telluric lines. We used the normalised spectra in our radial velocity measurements. 

For the SALT data we have adopted the results of the standard HRS pipeline for the spectra taken after commissioning of HRS was completed in late 2013 (80\% of the sample). The spectra obtained during commissioning (8 science observations in total) were reduced with the standard \textsc{iraf}\footnote{\textsc{iraf} (Image Reduction and Analysis Facility) is distributed by the National Optical Astronomy Observatories, which are operated by the Association of Universities for Research in Astronomy, Inc., under contract with the National Science Foundation.} reduction scripts from the \textsc{echelle} package. These shared risk spectra had to be addressed individually and were not well suited for reduction with the preliminary HRS pipeline at that time.

\subsection{Radial velocity determination}

We determined radial velocities using the \textsc{iraf} \textsc{fxcor} package. This package Fourier cross-correlates the observed spectrum with a template spectrum, measuring the relative shift between the two spectra. For each of the stars, we created template spectra using the MARCS~(Model Atmospheres in Radiative and
Convective Scheme) stellar atmospheres and the Turbospectrum spectral synthesis code (\citealt{Alvarez98}; \citealt{Gustafsson08}; \citealt{Plez08}) with stellar parameters T$_\mathrm{eff}$, $\log g$ and [Fe/H] as given in Table~\ref{table:programstars}. In \textsc{fxcor}, we fitted a Gaussian to the cross-correlation peak, from which the formal radial velocity uncertainties are determined following the method described in \citet{TonryDavis79}.

In stars that are so metal-poor as those in our sample, not many lines are present in the spectrum. There are only a few features present that can be used to determine robust radial velocities, the main features being the calcium triplet (at 8498, 8542 and 8662~\AA), the H$\alpha$ line (6563~\AA) and the magnesium triplet (5167, 5173 and 5184~\AA). Bluewards of these features the signal-to-noise ratios tend to be too low for good radial velocity measurements. 

\subsubsection{CFHT sample}
\begin{figure}
\centering
\includegraphics[width=\hsize,trim={0.0cm 0.6cm 0.0cm 0.0cm}]{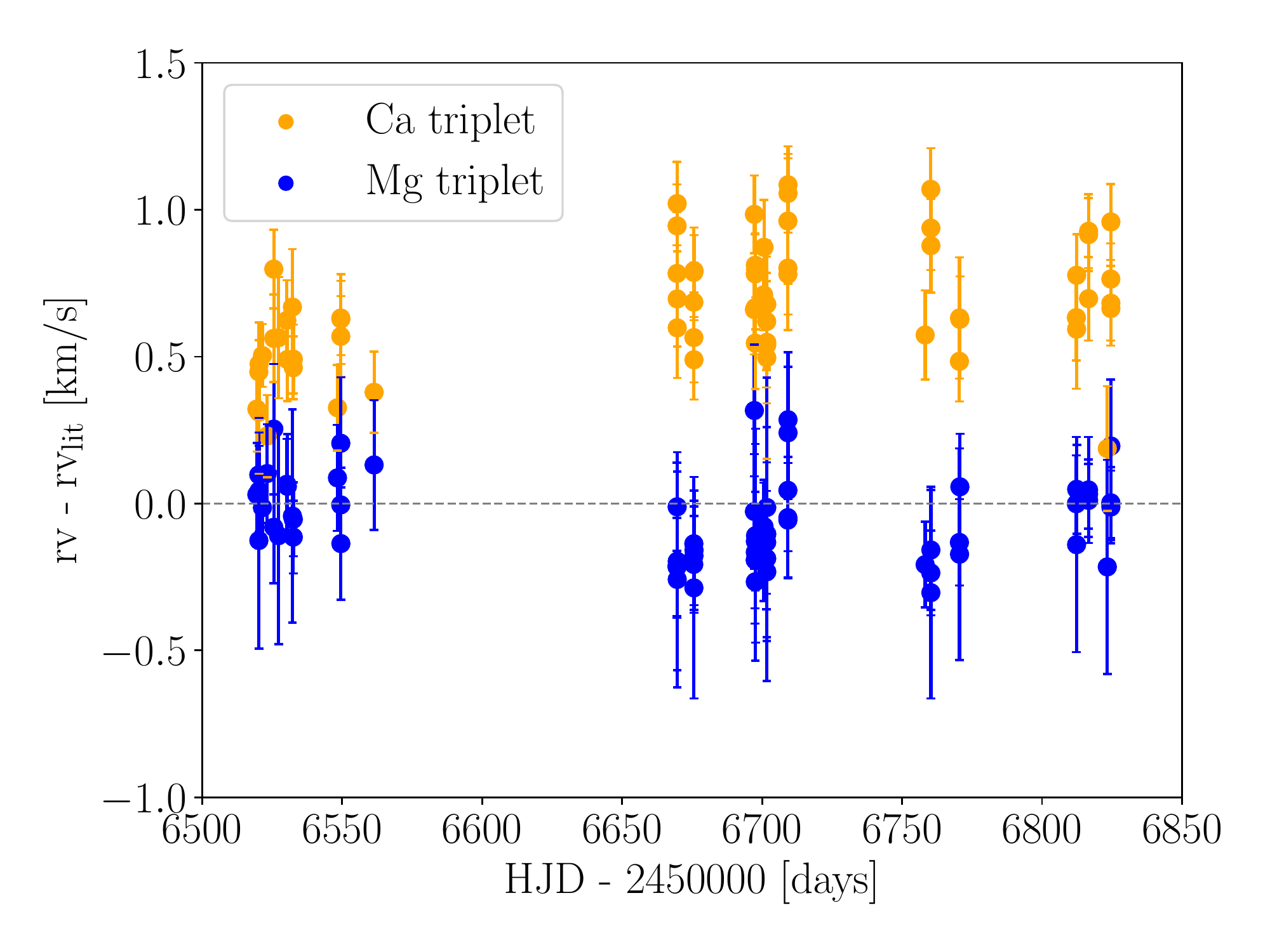}
\caption{Radial velocities of standard stars observed with CFHT with respect to their literature values. The expected zero-line is indicated.}
 \label{fig:standards}
\end{figure}

For our CFHT sample, we used the magnesium triplet region to determine good radial velocities. We found that the H$\alpha$ line is too broad for precise radial velocities, and when we used the calcium triplet we found a slight offset ($\sim 0.8$~\kms) for the standard stars with respect to the literature (see Figure~\ref{fig:standards}). From this figure we see that although the formal uncertainties on the radial velocities derived from the magnesium triplet are larger than those from the calcium triplet, the velocities are more accurate. Additionally, for the CEMP-no stars that overlap with the H16a sample the magnesium triplet radial velocities agree better with the H16a velocities than those from the calcium triplet. Our supposition is that the difference between the calcium triplet and magnesium triplet radial velocities is the result of a faulty wavelength calibration in the red part of the ESPaDOnS spectra. Therefore we decided to use the magnesium triplet radial velocities in our analysis for the CFHT spectra, with uncertainties as provided by \textsc{fxcor}.

\begin{figure}
\centering
\includegraphics[width=\hsize,trim={0.0cm 0.6cm 0.0cm 0.0cm}]{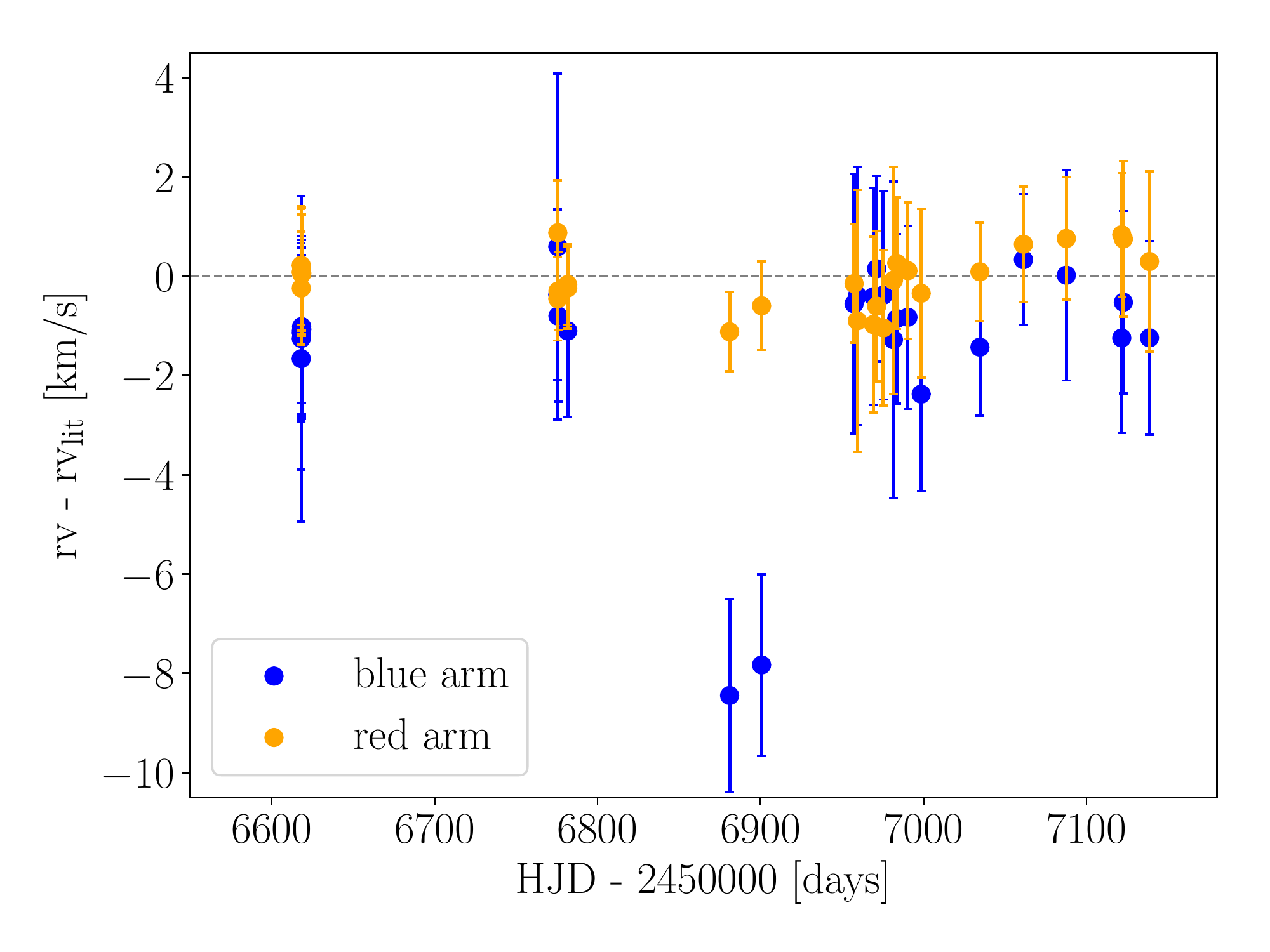}
\caption{Radial velocities of standard stars observed with SALT with respect to their literature values. The expected zero-line is indicated.}
 \label{fig:standardsSALT}
\end{figure}

\subsubsection{SALT sample}
The SALT spectra have lower signal-to-noise ratios than the CFHT spectra and were taken at lower resolution. For these observations we therefore used the spectra both from the red and blue arms to get more precise radial velocities. 

To correct for any instabilities in the instrument we computed two radial velocity corrections. First of all, we compute a correction from the telluric lines present in the red spectrum. Telluric lines have fixed wavelengths and can be used to correct exposure-to-exposure differences in the instrument which may cause changes in the wavelength solution. We apply this correction to both the blue and red spectrum of that observation, since there are no telluric lines present in the blue spectra. For the observed radial velocity standard stars, we present the telluric corrected radial velocities compared to the literature velocity in Figure~\ref{fig:standardsSALT}. The red arm velocities agree well with the literature within the uncertainties of about 1~\kms~for all standard stars across the full timespan, demonstrating that our radial velocity determination, when combined with the telluric correction, is robust and that the uncertainties are realistic.

Regarding the blue arm velocities for the standard stars, there is a small offset in most of the measurements. Furthermore, there is an especially large blue-red difference around HJD$\, -\, 2450000 = 6900$ days (August 2014) for the two observed standard stars. The history of maintenance operations on HRS shows that August 2014 was a time when there were several issues with the instrument. The vacuum was lost a few times during the first weeks of August, and afterwards the cameras needed to be heated up by more than 150 degrees to remove possible contaminants. Since this was done at least twice in this period, it is possible that in the process one of the CCDs moved slightly. This could lead to a discrepancy between the two arms.

However, we can correct the blue arm differences using the standard star observations. We apply a correction to the blue arm radial velocities for our science observations, which is the difference between the literature and measured radial velocity from the blue arm of the standard star(s) observed on the same night. Especially for the two science stars observed in August 2014 this drastically improves the consistency between the blue and red arm radial velocities. During a night, the blue arm correction may vary up to 1~\kms, and unfortunately not every science observation always has its own radial velocity standard observation. Therefore, we inflate the \textsc{fxcor} uncertainties on the blue arm radial velocities of the science observations by 1~\kms. 

The final radial velocities are computed as the weighted average of the (telluric and standard corrected) red and blue velocities, where the weights are the uncertainties as provided by \textsc{fxcor} (with the blue uncertainties inflated). Final radial velocity uncertainties are estimated by the standard deviation of the two velocities derived from the blue and red arm.

For almost half of the spectra there was no radial velocity standard star observed on the same night. For these stars we did not use the blue arm radial velocities, instead we accepted the radial velocities from the red arm with uncertainties as provided by \textsc{fxcor}. This resulted in lower precision for these stars, but as illustrated in Figure~\ref{fig:standardsSALT}, we are confident that the red arm velocities are accurate. Additionally, for two nights (five spectra) there was no red spectrum available for neither science nor standard stars, and therefore also no telluric correction. We exclude these measurements from this work entirely.

For the five stars with SALT data that overlap with the H16a sample, derived radial velocities agree with those from H16a within the uncertainties. 

\subsection{CEMP compilation}

For an overview of known CEMP stars, we have used as a baseline the work from \citet{Yoon16} who have compiled a large number of CEMP stars from the literature. It contains 127 CEMP-no stars, 147 CEMP-$s$ stars and 31 unclassified CEMP stars based on upper limits on their [Ba/Fe]. The carbon abundances in this compilation have been corrected for evolutionary phase (following \citealt{Placco14}), and throughout this paper we use their corrected [C/Fe] and A(C) values for the stars in our sample (Table~\ref{table:programstars}) and whenever we refer to the compilation. 

We have added six stars to this compilation: the recently discovered hyper metal-poor star SDSS~J0815$+$4729 from \citet{Aguado18} that is extremely enhanced in carbon (with [Fe/H] $< -5.8$ and [C/Fe] $> +5.0$), the recently discovered CEMP-no binary star SDSS~J1341$+$4741 from \citet{Bandyopadhyay18}, CS~22166$-$016 \citep{Giridhar01} and CS~22878$-$027 \citep{Yong13}, which are two CEMP-no stars monitored in radial velocity by H16a and us not present in the compilation, and G64$-$12 and G64$-$37 which were found to be CEMP-no stars by \citet{Placco16a}. For SDSS~J0929$+$0238 we have updated the $\log g$ (to the main-sequence solution), [Fe/H], [C/Fe], A(C), upper limit for [Ba/Fe] and binary status to the values in \citet{Caffau16}. We update the [Fe/H], [C/Fe] and A(C) of HE~2319$-$5228 to the values from \citet{Beers17}. HD~135148 is a known binary star \citep{Carney03}, so we updated its binary status in the compilation. The star has a [Ba/Fe] $= +0.3$ \citep{Simmerer04} and is not clearly classified as CEMP-no or CEMP-s, therefore we will not consider this star in the analysis of this work. For SMSS~0313$-$6708 we updated the [Fe/H] upper limit to $-6.5$ \citep{Nordlander17} .

For HE~1201$-$1512 and SDSS~J0140$+$2344, \citet{Yoon16} only provide the subgiant solutions, even though \citet{Yong13} provide both dwarf and subgiant solutions since the evolutionary status of this star was unknown at the time. With \textit{Gaia} DR2 \citep{Gaia18}, we can determine which solution is likely the best one. We compare BP$-$RP and absolute G magnitude (converted using the parallax) to a MIST isochrone\footnote{\url{http://waps.cfa.harvard.edu/MIST/interp_isos.html}} (\citealt{Dotter16}; \citealt{Choi16}) with [Fe/H] $= -4.0$ at an age of 12.5 Gyr. We find that both stars are more consistent with being dwarfs, therefore we accept the dwarf solutions.

\section{Results}

\label{sec:results}

\subsection{Radial velocity database}\label{sec:rvdatabase}

We present all derived radial velocities for the stars in our CFHT/SALT sample in Table~\ref{table:rvs}. 

We supplement our radial velocities with values from the literature, to get as large a timespan and as many radial velocities for as many stars as possible. We have compiled a list with all available radial velocity measurements from S14, H16a and this work, and added five stars from the literature. The CEMP-no stars G77$-$61 \citep{Dearborn86}, SDSS~J0929+0238 \citep{Caffau16} and SDSS~J1341$+$4741 \citep{Bandyopadhyay18} are stars known to vary in radial velocity. G77$-$61 and SDSS~J0929+0238 only have upper limits for [Ba/Fe], but are assumed to be CEMP-no stars because they are on the low carbon band. SDSS~J0929+0238 has been discovered as a double-lined (possibly even triple-lined) spectroscopic binary, and it was monitored for radial velocity variations after. For this star, two or three radial velocities per spectrum are given for spectra that had multiple components (as in \citealt{Caffau16}). The two stars G64$-$12 and G64$-$37 are CEMP-no stars that are constant in radial velocity \citep{Latham02}. Additionally, we kindly received several unpublished radial velocity measurements from N. Christlieb for HE~0557$-$4840 and HE~0107$-$5240, which we have added to the compilation. 

We supplemented our literature sample by deriving a radial velocity for other available spectra, using the same method as for the CFHT spectra. For HE~1201$-$1512, we derived the radial velocity from a FEROS spectrum that was taken during follow-up efforts from the Pristine survey (\citealt{Starkenburg17}; \citealt{Caffau17}). We also searched the ESO archive and found a UVES spectrum for SDSS~J0140+2344 taken as part of the TOPoS survey \citep{Bonifacio18} and 34 UVES spectra over the course of one year for the most iron-poor star SMSS~J0313$-$6708, which was observed as part of the SkyMapper extremely metal-poor star survey (\citealt{Keller14}; \citealt{Bessell15}). No radial velocities for SMSS~J0313$-$6708 have previously been published. 

Our efforts result in a sample of 710 individual radial velocity measurements (including this work) for 45 CEMP-no stars. The stars are listed in Table~\ref{table:rvcemplist} with their radial velocity properties, stellar parameters and carbon and barium abundances. The individual radial velocity measurements can be found in Table~\ref{table:litrvs}. 

\subsection{Radial velocity variation in the sample}

For each of the 45 CEMP-no stars in the compilation, we determine the $\chi^2$ of the radial velocity distribution,
\begin{equation} \label{eqn:X2}
    \chi^2 = \sum\limits_{i=1}^n \left( \frac{v_i - \bar{v}}{\sigma_{v_i}} \right)^2,
\end{equation} 

\noindent and use it to compute the probability that the radial velocity is constant, the p-value $P(\chi^2)$. Before computing the $\chi^2$, we quadratically add to the radial velocity uncertainties from H16a a floor uncertainty of 0.1~\kms~ to account for external uncertainty sources, as H16 did when computing their $\chi^2$. The final probability for each of the stars is presented in the fifth column of Table~\ref{table:rvcemplist}. \citet{Carney03} find that all binary stars in their sample have $P(\chi^2) < 10^{-6}$, which is what we take as our binary candidate selection criterion.

Among the stars with $P(\chi^2) < 10^{-6}$, we find the six known binary systems from the literature: CS~22957$-$027, HE~0219$-$1739, HE~1150$-$0428 and HE~1506$-$0113, which are the four binaries discussed in S14 and H16a, and additionally G77$-$61 and SDSS~J1341$+$4741 (see Section~\ref{sec:rvdatabase}). SDSS~J0929+0238 is also binary but not included in this analysis since it is a spectroscopic double-lined system.

\subsubsection{New binary candidates}

There are five additional stars with $P(\chi^2) < 10^{-6}$, which are good binary candidates. Three of these, HE~0107$-$5240, HE~2139$-$5432 and SDSS~J1422$+$0031, are in our southern hemisphere SALT sample. The fourth star, SDSS~J0140$+$2344, is in our CFHT sample and is one of the few stars in that sample that has not been monitored by H16a. Finally SDSS~J1313$-$0019 has not been observed by us but comes from the literature. 

\begin{figure*}
\centering
\includegraphics[width=\textwidth,trim={3.0cm 0.1cm 2.7cm 0.0cm}]{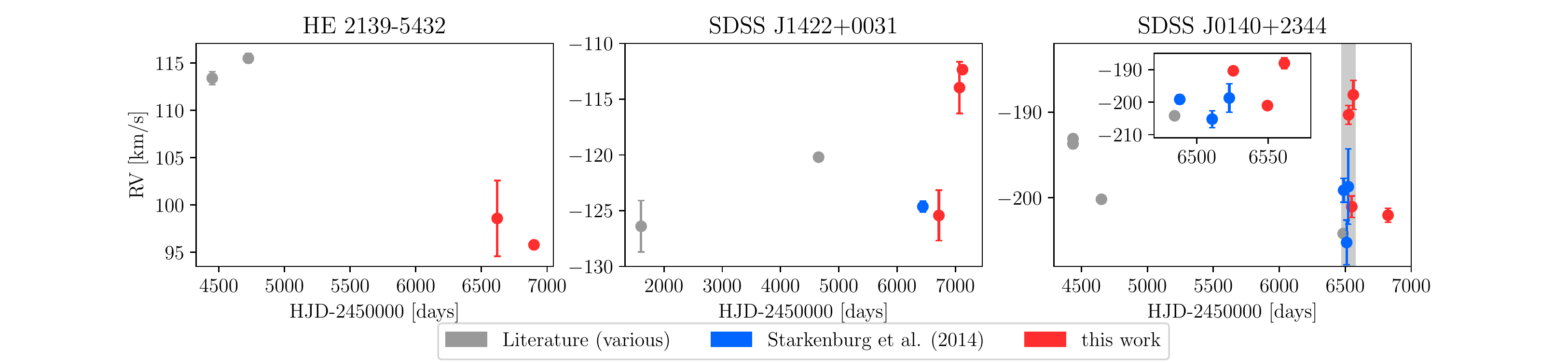}
\caption{Radial velocities for the three stars in our sample varying strongly in radial velocity ($> 10$~\kms) that were not previously identified as binaries. For SDSS~J0140+2344, a zoom-in of the shaded region is shown. }
 \label{fig:binaries}
\end{figure*}

We present the individual radial velocity measurements for the three stars with $> 10$~\kms~variation that include measurements from this work in Figure~\ref{fig:binaries}. SDSS~J1422$+$0031 and SDSS~J0140$+$2344 the latter two were already mentioned as possible interesting candidates in S14.

\begin{figure}
\centering
\includegraphics[width=\hsize,trim={0.0cm 0.9cm 0.0cm 0.0cm}]{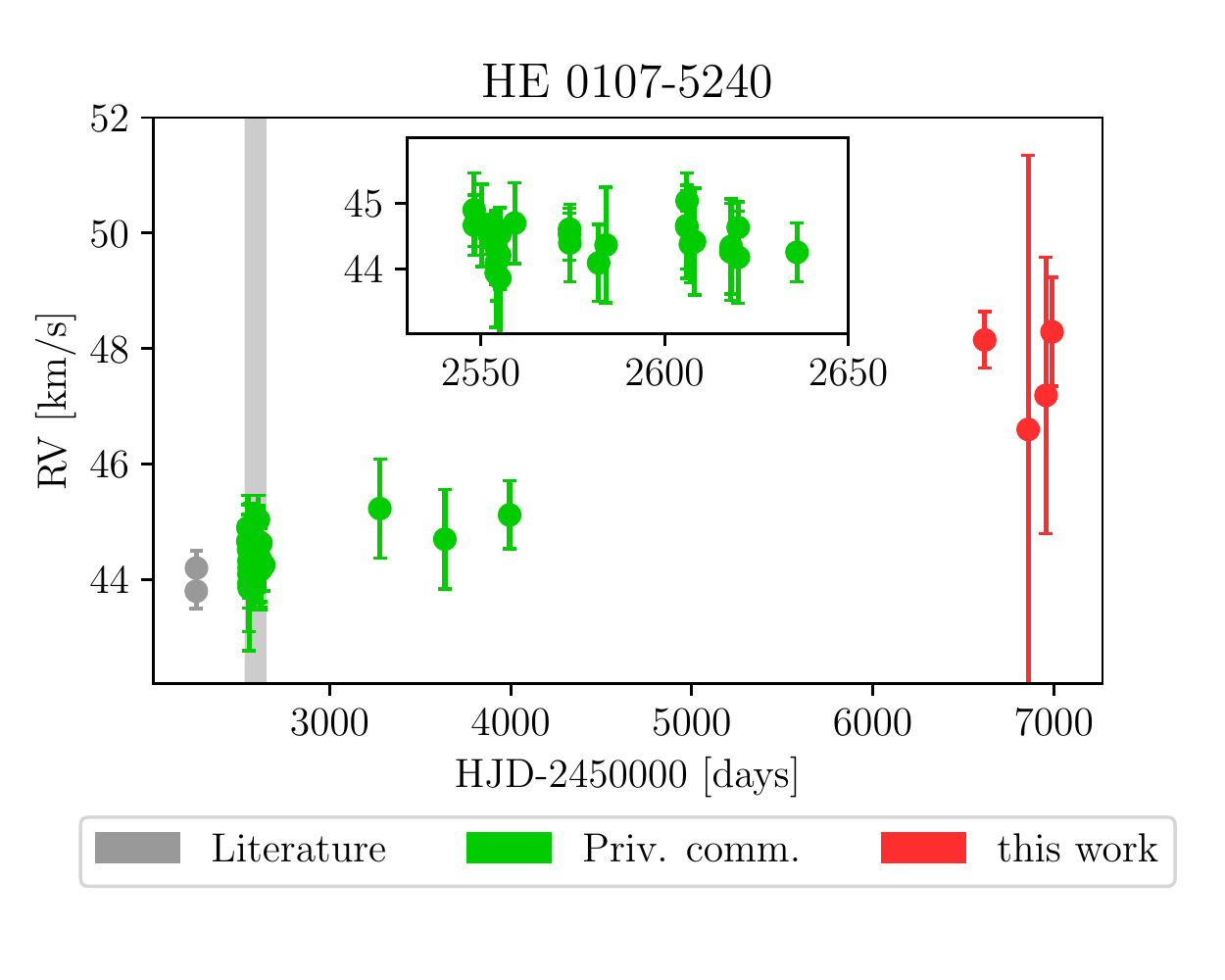}
\caption{Radial velocities for HE~0107$-$5240, including those unpublished values which we received via private communication from N. Christlieb. A zoom-in of the shaded region is shown.}
 \label{fig:HE0107}
\end{figure}

The fourth star that includes measurements from this work, HE~0107$-$5240, is presented in Figure~\ref{fig:HE0107}. This star was long thought to be non-variable in radial velocity, however, when including our new measurements this star appears to be varying on a large timescale ($>10$ years). Our supposition is that it is part of a (wide) binary system. 

The final star, SDSS~J1313$-$0019, has three radial velocity measurements in the literature, of which two come from low-resolution spectra (268 $\pm$ 4 \kms~and 242 $\pm$ 4 \kms~from SEGUE and BOSS respectively, \citealt{Allende15}) and one from a high-resolution spectrum (274.6 \kms, no uncertainty given, \citealt{Frebel15}). Both authors have suggested this star might be in a binary system, but more measurements are needed to confirm. We do not treat this star as a binary system in this work because it has only one measurement from high-resolution spectroscopy. 

\subsubsection{Other stars with low $P(\chi^2)$}

There were two additional stars that had $P(\chi^2) < 10^{-6}$. The first, HE~1410+0213, was observed extensively by H16a and after much analysis they concluded that the star is most likely single. They assume that the velocity variation comes from low-amplitude pulsations in the star and suggest adding a velocity jitter of 0.15~\kms. When we add such a jitter, $P(\chi^2)$ is increased (as presented in Table~\ref{table:rvcemplist}). 

The second star is the most iron-poor star of the compilation, SMSS~0313$-$6708. Its 34 radial velocities (which we derived from archive UVES spectra) measured over the course of one year have a dispersion of 0.4~\kms, with two measurements that have difference of 2.3~\kms~(see Figure~\ref{fig:SMSS0313} in the Appendix). We were not able to fit an orbit through all points using the method in Section~\ref{sec:orbits} and, excluding the two outliers, the other measurements seem compatible with a constant radial velocity. We therefore conclude that it is likely that the radial velocity uncertainties are underestimated. In the ESO archive, there are no radial velocity standard stars that are observed on the same nights as the two outliers, therefore we cannot correct for any systematics. Instead, we estimate an uncertainty floor by assuming that the star is not varying in radial velocity. We quadratically add a constant to each of the the uncertainties until a final reduced $\chi^2$ of 1 is reached. This leads to an uncertainty floor of 0.35~\kms, which we have added to each measurement. These inflated uncertainties are what we provide for SMSS~0313$-$6708 in Table~\ref{table:litrvs}. After this correction, this star does not have $P(\chi^2) < 10^{-6}$ anymore (the updated value is presented in Table~\ref{table:rvcemplist}). In this work we treat SMSS~0313$-$6708 as single, although it could still vary in radial velocity on a longer timescale than one year. 

\subsubsection{Final notes on radial velocity variations}

Important to note is that radial velocity variations are not only caused by binarity, they can also be caused for example by inhomogeneities on the surfaces of stars or stellar pulsations. \citet{Carney08} find that the velocity ``jitter" due to inhomogeneities is mainly affecting evolved stars with $M_V \leq -1.4$ ($\log g \lesssim 1.0$). None of our four new binary candidates are in this regime, nor in parts of the HR diagram known for stellar pulsations, therefore we conclude that their radial velocity variations are due to binarity. 

For all remaining stars in our CFHT/SALT sample we present the individual radial velocity measurements in Figure~\ref{fig:appendixrvs}. Our  measurements agree with previous measurements wherever there is overlap. 

\begin{figure*}
\centering
\includegraphics[width=\textwidth,trim={1.0cm 0.4cm 0.5cm 1.0cm}]{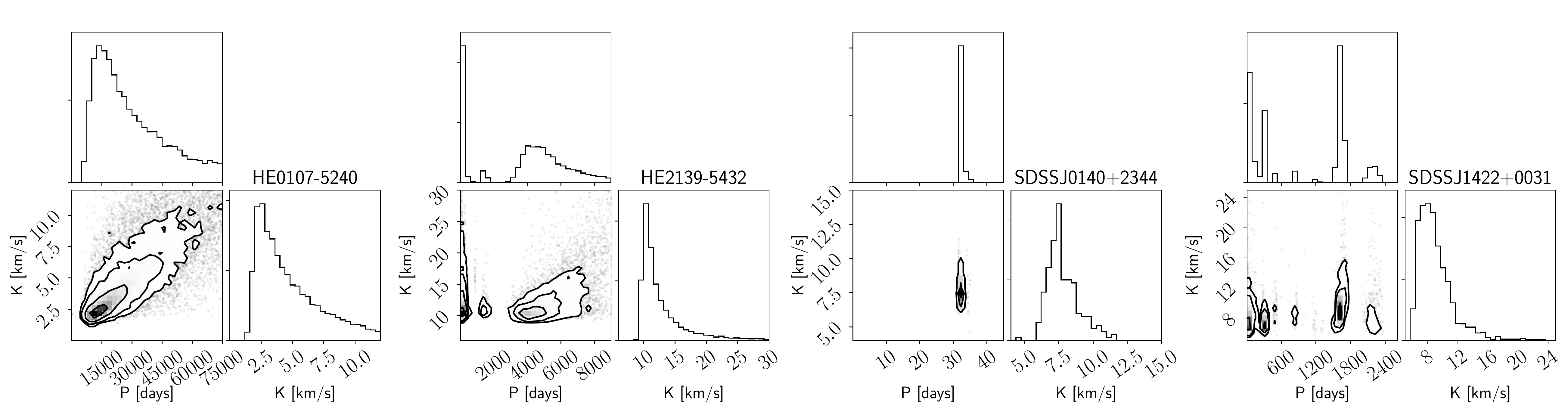}
\caption{Orbits solutions from \textsc{the Joker} for the period P and the semi-amplitude K for the new binaries. The ranges on the x- and y-axes have been truncated for clarity, however at least the two largest peaks of the distribution are always shown. }
 \label{fig:orbits}
\end{figure*}

\subsection{Orbit properties of the new binaries}\label{sec:orbits}

We apply the code \textsc{the Joker} (\citealt{PW17a}; \citealt{PW17b}) to the radial velocity data of our four newly discovered binary systems: HE~0107$-$5240, HE~2139$-$5432, SDSS~J0140$+$2344 and SDSS~J1422$+$0031. 
\textsc{The Joker} is a Monte-Carlo sampler for orbital parameters of binary systems that can also be applied to sparse and/or low-quality radial velocity data. It produces a posterior sampling of the period, eccentricity, pericentre phase and argument, velocity semi-amplitude and the barycentre velocity. For all four of our stars, the eccentricity and pericentre phase and argument were not well-constrained in the analysis. The resulting corner plots \citep{Foreman16} for the periods and semi-amplitudes however are insightful and can be found in Figure~\ref{fig:orbits}. We took $10^6$ samples per star, except for SDSS~J0140$+$2344 where we took $10^8$ because the orbit is relatively well-determined with the available radial velocity measurements, so few samples will be accepted.

The analysis with \textsc{the Joker} for HE~0107$-$5240 results in a period distribution that peaks between 10000 and 30000 days ($27 - 82$ years) and a semi-amplitude of the order of 2--5~\kms. HE~2139$-$5432 has sparse radial velocity data that allow for many possible short-period orbits up to 300 days, or longer period orbits of $\sim$4000 days, both with semi-amplitudes of $\sim$11~\kms. For SDSS~J0140$+$2344 we find a narrow peak of the period distribution at 32 days, and a semi-amplitude of 7.5~\kms. Finally, for SDSS~J1422$+$0031 we find multiple peaks, the most pronounced one producing a period of $\sim$1600 days and a semi-amplitude of $\sim$8~\kms. Clearly for at least three of these stars more radial velocity measurements are needed to determine better orbital solutions.

\section{Properties of the CEMP-no radial velocity sample}
\label{sec:binarysample}

By combining our four new CEMP-no binaries with the four known binaries discussed in S14 and H16a (CS~22957$-$027, HE~0219$-$1739, HE~1150$-$0428 and HE1506$-$0113) and the three literature stars G77$-$61 \citep{Dearborn86}, SDSS~J0929+0238 \citep{Caffau16} and SDSS~J1341$+$4741 \citep{Bandyopadhyay18}, we have a sample of eleven CEMP-no binary stars. 

To get to a binary fraction, the number of single stars should also be determined. Long monitoring time-scales and high radial velocity precision are needed to rule out the binarity of a star. If the time-scale is too short or the uncertainties are too large, a long-period low-amplitude signal could possibly be hiding in the data. For all practical purposes however we will assume that the monitored stars that do not show indications of radial velocity variations are single. We exclude from the single star sample those stars with fewer than five radial velocity measurements, because the radial velocity precision and/or temporal coverage of those measurements is not sufficient to claim that a star is single based on so few data points. Combining the single and binary stars results in a binary fraction of $32^{+10\,}_{\,\,-9}\%$ (11 out of 34) for the whole sample of radial velocity monitored CEMP-no stars, using binomial statistics to derive the $\pm1\sigma$ uncertainties. This binary fraction is larger than $17^{\,+11\,}_{\,\,-8} \%$ (4 out of 24) as found by H16a for CEMP-no stars, but agrees within 1$\sigma$. The discrepancy of our result with the binary fraction of $16^{+5\,}_{-4} \%$ (14 out of 85) for carbon-normal metal-poor ($ -3.0\lesssim \mathrm{[Fe/H]} \lesssim -1.4$) giants by \citet{Carney03} is slightly larger than 1$\sigma$. However, both of these are not yet statistically significant differences. 

The sample of H16a appears to be more homogeneously monitored than the complete combined sample in this work, because we and others in the literature may have preferentially monitored stars that already showed some indication of radial velocity variation. The binary fraction in this work might not necessarily be representative for the whole population.

Actually, all these binary fractions (including those from H16a and \citealt{Carney03}) should better be thought of as lower limits, given that long-period binaries or binaries at large inclinations may still be present among the assumed single stars. It is much easier to confirm the binarity of a star than to rule out its binarity, since fewer measurements are needed to find that a star is variable in radial velocity, especially if the amplitude of the variation is large.  

However, the collection of binary stars we now have is interesting to study in itself. In the following subsections we discuss some of the properties of the CEMP-no binary population.

\begin{figure}
\centering
\includegraphics[width=\hsize,trim={0.2cm 0.4cm 0.0cm 0.0cm}]{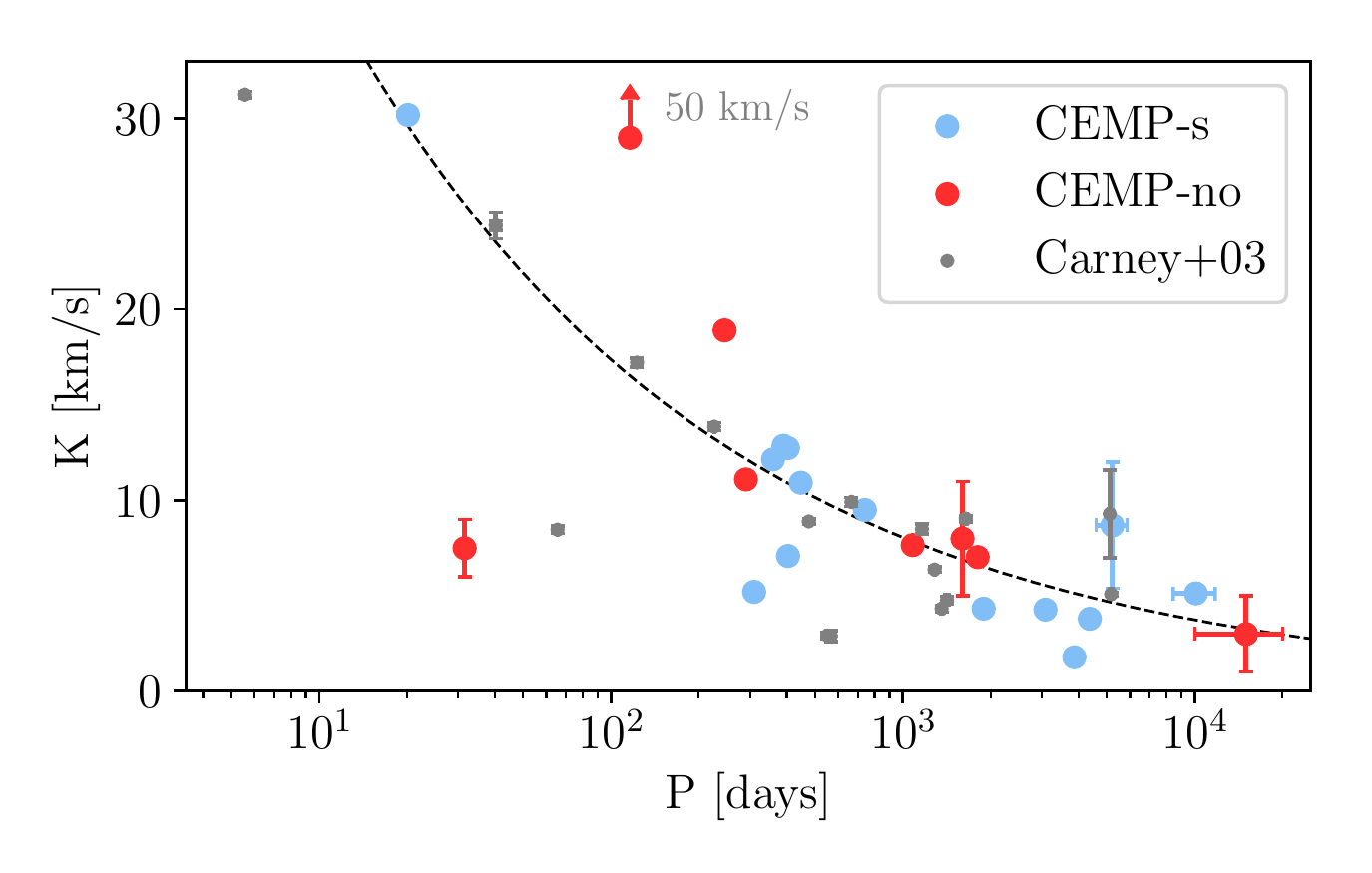}
\caption{Periods and semi-amplitudes of carbon-normal metal-poor binaries from \citet{Carney03}, CEMP-$s$ binaries from \citet{Hansen16b} (excluding their CEMP-r/s stars) and CEMP-no binaries (compilation from this work). HE~1506$-$0113 (see Figure~\ref{fig:appendixrvs}) and SDSS~J0929$+$0238 are not shown on this figure since they do not have any derived orbits, HE2139-5432 is not shown because it either has P $< 300$ or P $\approx$ 4000 days (with K $\approx 11$ \kms). The line indicates what would be expected of Keplerian orbits of a 0.8 M$_\odot$ star with a 0.5 M$_\odot$ companion, for an eccentricity of 0.3 (typical for the stars in the \citealt{Carney03} sample) under an inclination of 60$^\circ$.}
 \label{fig:PK}
\end{figure}

\begin{figure*}
\centering
\includegraphics[width=\textwidth,trim={2.0cm 0.0cm 2.5cm 0.0cm}]{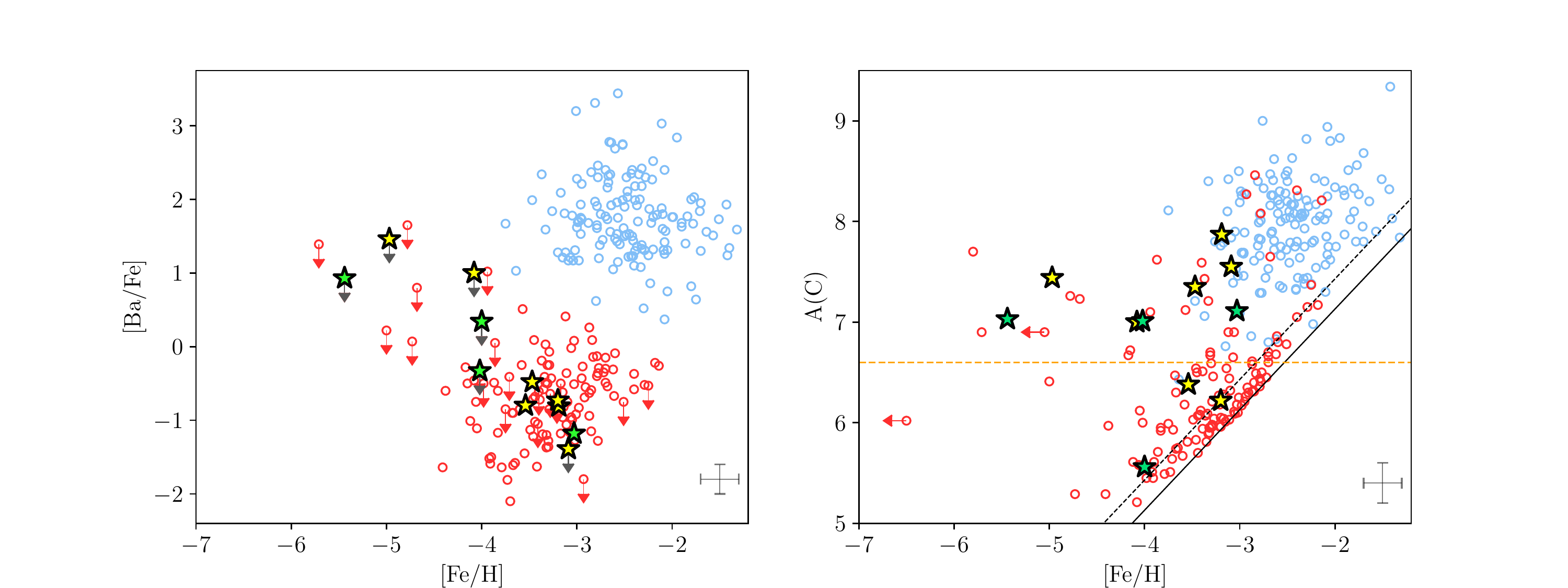}
\caption{Left: [Ba/Fe] as a function of [Fe/H] for the compilation of CEMP stars from \citet{Yoon16} down to [Fe/H] $= -6$, where CEMP-$s$ stars are shown in blue and the CEMP-no stars in red. Indicated are the known and new CEMP-no binaries as yellow and green stars respectively. Right: The same but for A(C) on the y-axis. The solid and dotted black lines indicate the [C/Fe] $> +0.7$ and [C/Fe] $> +1.0$ CEMP criteria respectively. The orange dashed line indicates the separation of the low and the intermediate/high carbon band. For both panels, typical error bars on the abundances are shown in the lower right corner.}
 \label{fig:combi}
\end{figure*}

\subsection{Orbit characteristics} \label{subsec:orbits}

How do the orbit properties of CEMP-no stars compare to those of CEMP-$s$ stars and other metal-poor binary systems? We compare the periods and semi-amplitudes in Figure~\ref{fig:PK}. The uncertainties for the binaries from this work are taken from the $1 \sigma$ contours of the most significant peaks from the probability distributions in Figure~\ref{fig:orbits}. There is no indication that the CEMP-no binaries are of a different distribution than the CEMP-$s$ binaries or the stars from \citet{Carney03}. The typical companion mass for stars in the latter sample is 0.5 M$_\odot$ with eccentricities of $\sim 0.3$. Only three of the  CEMP-no binaries have known eccentricities, H16a claim that their distribution is not different from normal metal-poor stars. 

One odd star is SDSS~J0140$+$2344 with its short period of 31.5 days and a relatively low semi-amplitude of 7.5~\kms. It is a possibility that this system is observed relatively face-on. Another odd star is SDSS~J1341$+$4741, which has a reported period of 116 days and a semi-amplitude of 50 \kms~\citep{Bandyopadhyay18}. If this is confirmed with more radial velocity measurements, this star would have to have a more massive companion than the other stars or a very eccentric orbit.

H16a could not derive an orbital solution for HE~1506$-$0113, despite the large number of radial velocity measurements and its clear variability. This star seems to vary on a small time-scale ($\sim$~20 days) in the data from \citet{Norris13a} that have been reanalysed by H16a, and a larger time-scale ($\sim$~1000 days) on the basis of data from S14 and H16a, see Figure~\ref{fig:appendixrvs}. Our new radial velocities for this star fill the gap between the measurements of S14 and H16a, but do not help to clarify its orbit. 

\subsection{Enhancement in s-process elements}
Enhancement in the s-process element barium is usually a sign of mass transfer from an AGB companion. We present [Ba/Fe] as a function of [Fe/H] in the left panel of Figure~\ref{fig:combi} for our binary stars on top of the CEMP compilation of \citet{Yoon16}, where the CEMP-no stars are shown in red and the CEMP-$s$ stars in blue. The CEMP-no binaries from the literature (from S14, H16a, \citealt{Dearborn86}, \citealt{Caffau16} and \citealt{Bandyopadhyay18}) are shown as yellow stars, and the new binaries uncovered in this work as green stars. CEMP star classes are defined mainly by the barium abundance, therefore the CEMP-$s$ stars (blue points) and the CEMP-no stars (red points) separate almost perfectly in this diagram. There are four binaries with upper limits on [Ba/Fe] that are above zero. HE~0107$-$5240 and SDSS~J0140$+$2344 were already discussed in Section~\ref{sec:sample}, and taking the revised CEMP-no definition of \citet{Matsuno17} SDSS~J0929$+$0238 is also classified as CEMP-no. Even though it does not satisfy the revised definition, the dwarf G77$-$61 ([Ba/Fe] $< +1.0$) is likely a CEMP-no star too, assuming that all ultra metal-poor low-carbon band stars are. Alternatively, it could be the first ultra metal-poor ([Fe/H] $\leq -4.0$) CEMP-$s$ star. However, since no ultra metal-poor CEMP-$s$ stars are known to date, we assume that it belongs to the CEMP-no class. 

The binaries seem to be part of the normal CEMP-no distribution in the left panel of Figure~\ref{fig:combi}. The low [Ba/Fe] values (or upper limits) of most the binary stars are consistent with having had no ``classical" binary interaction with an AGB star in which s-process elements have been transferred together with carbon.   

\begin{figure}
\centering
\includegraphics[width=\hsize,trim={0.3cm 0.4cm 0.0cm 0.0cm}]{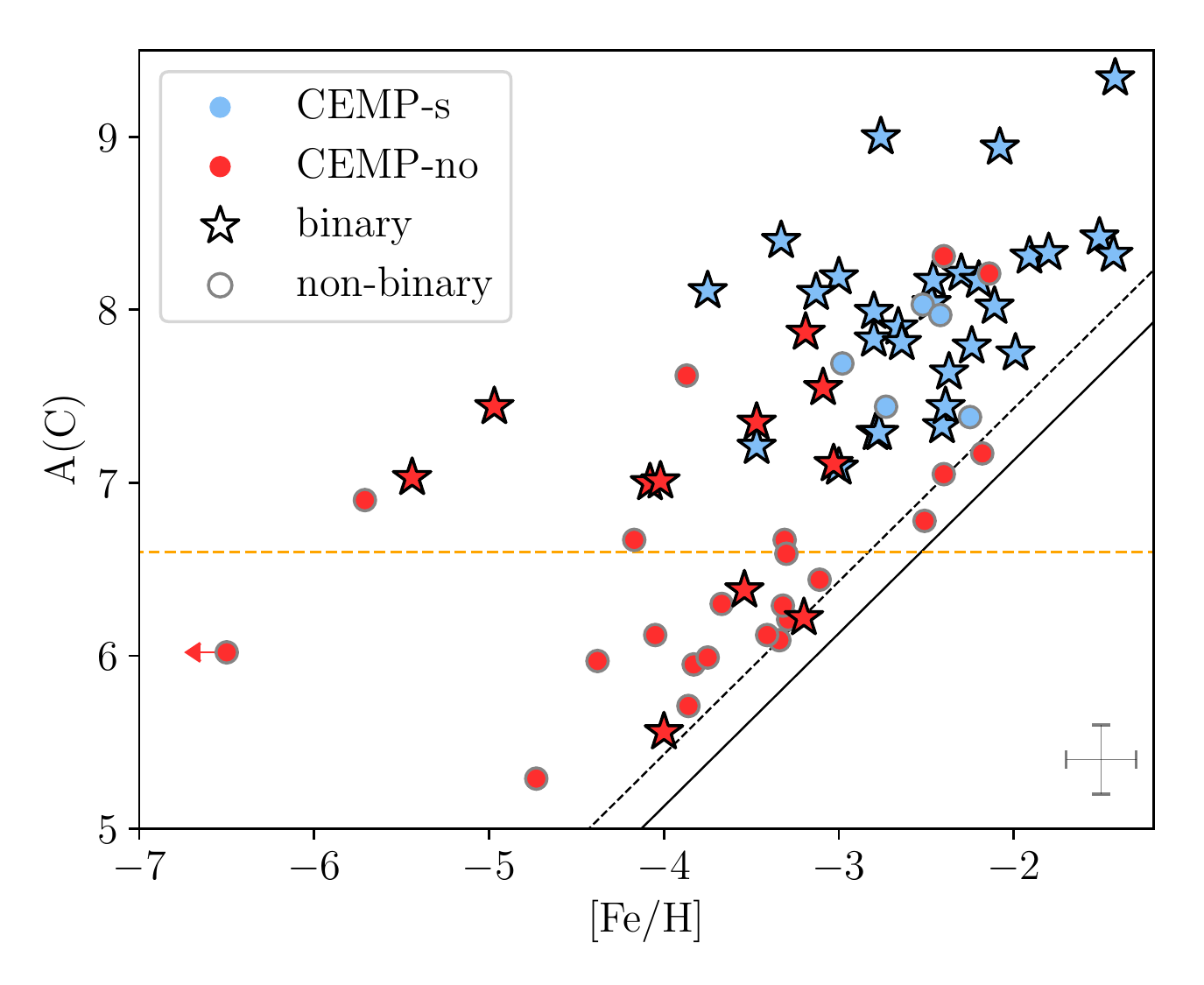}
\caption{A(C) as a function of [Fe/H] for CEMP stars, where again CEMP-no stars are shown in red and CEMP-$s$ stars in blue. Binary stars are indicated by a star symbol. The orange dashed line is the same as in the right panel of Figure~\ref{fig:combi}. }
 \label{fig:AC}
\end{figure}

\subsection{Absolute carbon abundance}
As discussed in the introduction, in general CEMP-$s$ and CEMP-no stars have different absolute carbon abundances. This can be seen in the right panel of Figure~\ref{fig:combi} for the compilation of CEMP stars. \citet{Yoon16} suggest that there may be three different groups of CEMP stars based on their A(C) and [Fe/H]. First, there are the Group I stars that cover the region of the CEMP-$s$ stars at large A(C) and higher [Fe/H]. Then there are two different groups of CEMP-no stars, the Group II stars that in general have lower A(C) and exhibit a clear dependence of A(C) on [Fe/H], and the Group III stars that seem to centre on a higher A(C) without a clear dependence on [Fe/H]. 

One might expect that like the CEMP-$s$ stars, most high carbon band CEMP-no stars would be in binary systems. However, H16a did not find a strong correlation between the binary status of CEMP-no stars and their location on the A(C) versus [Fe/H] plane (although that is difficult to say with such a small sample). In the right panel of Figure~\ref{fig:combi} we highlight the eleven currently known CEMP-no binaries on the A(C) versus [Fe/H] plane. It appears that most of the CEMP-no binaries have relatively high A(C) values in between the high and low carbon bands of \citet{Spite13}, which seems to correspond roughly to the region of the Group III stars of \citet{Yoon16} (although four of the binary stars are actually classified as Group I stars).

This does not seem to be a selection effect of only monitoring stars with the highest carbon enhancement, which can be seen from Figure~\ref{fig:AC}. There we present the same A(C) versus [Fe/H] plane as in Figure~\ref{fig:combi}, but now only for stars that have sufficient radial velocity data to say with some confidence whether they are in a binary system or not. Typically the stars that we assume to be single have been observed as thoroughly as the other stars, but we cannot fully exclude the possibility that they are in a binary system with a long period or low amplitude. For the CEMP-$s$ stars, we used the binarity information as documented in \citet{Yoon16}. 

In Figure~\ref{fig:AC} we notice that the fraction of CEMP-no stars that are in binary systems seems higher for stars on the intermediate/high carbon band compared to the stars on the low carbon band. Splitting our sample of well-monitored stars in half based on A(C) as illustrated by the orange dashed line in Figure~\ref{fig:AC}, we find that for the CEMP-no stars with A(C) $> 6.6$ the binary fraction is $47^{\,+15\,}_{\,-14}\%$ (8 out of 17), and for A(C) $\leq 6.6$ it is $18^{\,+14\,}_{\,\,-9} \%$ (3 out of 17). There is a 1$\sigma$ difference between these two fractions. 

If we conservatively assume that except for the known binary CEMP-no stars, all the known CEMP-no stars with A(C) $> 6.6$ are single stars (even if they do not have any radial velocity information), we find a binary fraction of $18^{\,+8\,}_{\,-6} \%$ (8 out of 44). This conservative lower limit for the binary fraction of high carbon CEMP-no stars is independent of the selection for radial velocity monitoring or the quality of the determination of single stars. 

The periods for CEMP-no binary stars with A(C) $> 6.6$ and derived orbits are similar to the periods for CEMP-$s$ binary stars, which typically are of the order of a few 100 to a few 1000 days (see Figure~\ref{fig:PK}). The three CEMP-no binaries with A(C) $< 6.6$ are HE~1506$-$0113, SDSS~J0140$+$2344 and SDSS~J1341$+$4741. It is curious that each of these three stars was described in Section~\ref{subsec:orbits} because they have no, or odd orbital solutions. 

We end this section with a note of caution: one should be careful when interpreting (absolute) carbon abundances, since most of the measurements were not done using non-LTE and/or 3D models, and such corrections may be especially important for carbon when comparing stars of different metallicities and evolutionary stages.

\begin{figure}
\centering
\includegraphics[width=\hsize,trim={0.3cm 0.4cm 0.0cm 0.0cm}]{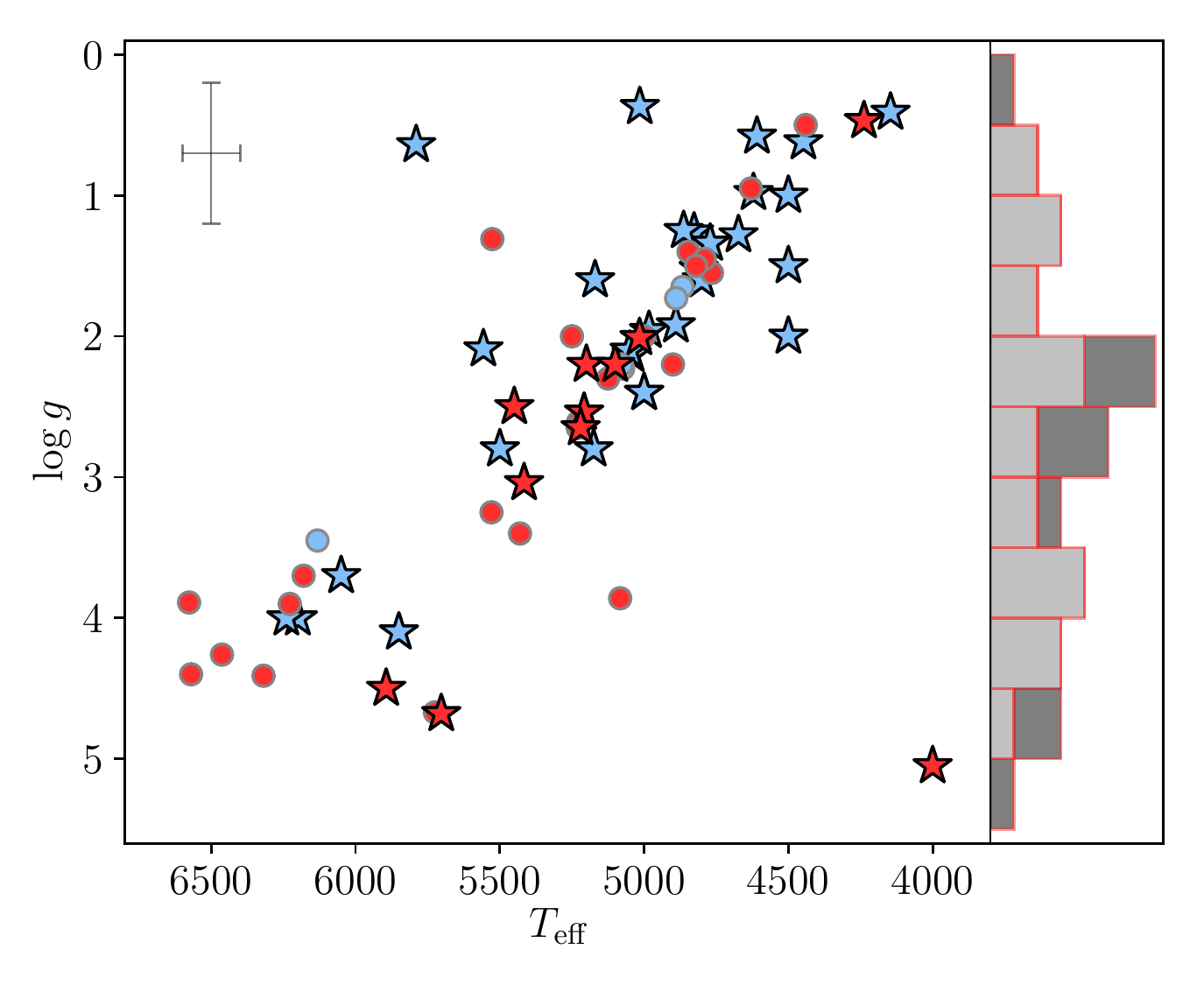}
\caption{Hertzsprung-Russell diagram for all CEMP-no and CEMP-$s$ stars, with the same symbols and colour coding as in Figure~\ref{fig:AC}. Because of the crowdedness on the narrow giant branch, for the CEMP-no stars we additionally show the stacked histogram for $\log g$ on the side, where dark grey are binary and light grey are single stars. Typical error bars are indicated in the top left corner.}
 \label{fig:HR}
\end{figure}

\subsection{Hertzsprung-Russell diagram}
We highlight the location of the CEMP stars on the (spectroscopic) Hertzsprung-Russell diagram in Figure~\ref{fig:HR}, where again we only show the stars that have sufficient radial velocity information available. Almost all the stars in the compilation are giant stars, since these are intrinsically brighter and easier to analyse. Additionally, CEMP stars with lower effective temperatures are easier to recognise from the spectra because the CH features are more distinct. There are three CEMP-no binary stars on the main sequence (or close to the turn-off), G77$-$61 with $T_\mathrm{eff} = 4000$~K, SDSS~J0140$+$2344 with $T_\mathrm{eff} = 5703$ and $\log g = 4.7$, and SDSS~J0929$+$0238 with $T_\mathrm{eff} = 5894$~K and $\log g = 4.5$. Another possibility is that SDSS~J0929$+$0238 is a sub-giant branch star with $\log g = 3.7$, but it is argued by \citet{Caffau16} that the main-sequence solution is more likely (unfortunately the \textit{Gaia} DR2 parallax for this star is uncertain). 

One of the CEMP-no binary stars, HE~0219$-$1739, is at the tip of the giant branch.\footnote{The stellar parameters for this star are however somewhat uncertain, T. Hansen found $T_\mathrm{eff} = 4600$ and $\log g = 2.3$ with stacked high-resolution radial-velocity monitoring spectra (priv. comm.). Unfortunately the \textit{Gaia }parallax for this star is negative and does not help to constrain the $\log g$.} The question arises whether this star is a real CEMP star, or alternatively an intrinsic (pulsating) carbon star. The well-determined radial velocity period of this star is 1800 days, which is longer than typical pulsation periods of long-period variable (LPV) stars. This star is also indicated as variable in its photometry by PanSTARRS \citep{Hernitschek16}, but the given variability time-scale (0.13 days) and magnitude amplitude (0.034 mag in r) are too small to represent LPV pulsations.  We conclude that this star is most likely a bona fide CEMP star in a binary system. 

It is curious that there seems to be an over-density of CEMP-no binary stars between $\log g$~=~2--3 as illustrated in a histogram in the right panel of Figure~\ref{fig:HR}, even though we have monitored stars along the entire giant branch and turn-off. In comparison, the general sample of CEMP-no stars (from the \citealt{Yoon16} compilation) does not show an over-density between $\log g$~=~2--3. Larger samples are needed to put this tentative result on firmer~footing. 

We note that compared to CEMP-no stars, there seems to be a larger number of monitored CEMP-s stars on the upper part of the red giant branch. It is unlikely that this is a brightness selection-effect, since the stars in the samples of H16a and \citet{Hansen16b} for CEMP-no and CEMP-$s$ stars respectively cover similar ranges in V magnitudes (and they make up most of the stars in this diagram). There could however be other unknown selection effects that play a role. 

\section{Discussion}
\label{sec:discussion}

\subsection{CEMP-no stars and binary mass transfer}

\label{sec:bintransfer}

Each of the CEMP-no binary stars must have a companion that causes its radial velocity variations. Since none of the stars except SDSS~J0929$+$0238 seem to be double-lined spectroscopic binaries and no visual counterpart is seen in photometry of any of the stars, the companion must be a fainter star. For the dwarf star G77$-$61 it has been argued that the companion is most likely a white dwarf \citep{Dearborn86}. For the giants, the companion can be expected to be a main-sequence star or a white dwarf (which has gone through the AGB phase in the past). Would it therefore not be likely that if some of the CEMP-no stars have white dwarf companions, they might have been polluted by mass transfer from an AGB companion at some point in their life? 

For CEMP-$s$ stars, mass transfer from an AGB star is the main formation scenario, which has been confirmed by the number of them that is in a binary system ($82 \pm 10 \%$, \citealt{Hansen16b}). Different types of mass-transfer are effective at different initial separations of the stars in the binary system, where Roche-lobe overflow can happen in the closest systems, wind-assisted Roche-lobe overflow in systems of intermediate separation and wind transfer by itself in wider systems. As shown in Figure~\ref{fig:PK}, typical periods and radial-velocity amplitudes for binary CEMP-$s$ and CEMP-no stars are similar. How many of the CEMP-no stars would have experienced binary interaction?

A simple estimate of the general fraction of metal-poor stars interacting with a binary companion can be made using the CEMP-$s$ stars, assuming that they all became CEMP-$s$ by mass-transfer from a former AGB star. Of all stars with [Fe/H] $< -2.0$, 13\% is a CEMP-$s$ star \citep{Placco14} and therefore in a formerly interacting binary system. There is no reason (yet) to assume that at lower metallicity, binary stars suddenly start interacting less. In fact, by re-examining different surveys, \citet{Moe18} have found that the close binary fraction (P < 10$^4$ days) of low mass stars (0.6-1.5 M$_{\odot}$) increases with decreasing metallicity. Therefore, in the regime of the CEMP-no stars, which are at lower metallicity than the CEMP-$s$ stars, one might expect an even larger fraction of all stars to have had interaction with a companion, simply because the binary fraction is higher. But we see almost no CEMP-$s$ stars with [Fe/H] $< -3.0$, and none at all with [Fe/H] $< -4.0$. So what do extremely metal-poor stars that underwent mass-transfer from a former extremely metal-poor AGB companion look like? 

Our CEMP-no binaries are not enhanced in s-process elements (specifically barium), which has usually been taken as a sign that they have not had any interaction with a binary companion. However, much is unknown about ultra metal-poor AGB stars (see the recent review by \citealt{Gil-pons18}). Models of AGB transfer among metal-poor stars have focussed on CEMP-$s$ stars (as in \citealt{Abate15}), which are mostly found at higher metallicities than the CEMP-no stars (see Figure~\ref{fig:combi}). The most metal-poor AGB yields available with s-process elements only go down to [Fe/H]~=~$-2.3$ \citep{Lugaro12}. It is potentially the case that extremely or even more metal-poor AGB stars produce fewer s-process elements, as for example in the (non-rotating) models of \citet{Suda04}, \citet{Lau07} and \citet{Cruz13}. Additionally, rotation can strongly affect the s-process element production, as for example in intermediate mass spinstars \citep{Meynet10}. Furthermore, the mass of the AGB star is important. It is expected that intermediate mass AGB stars  produce fewer s-process elements compared to AGB stars of lower mass, which especially affects s-process elements beyond the first s-process peak \citep{KarakasLattanzio14}. If these are the polluting companion stars, no barium excess should be expected. Finally, \citet{Busso99} suggested that in very metal-poor AGB stars, the $s$-process mainly produces third peak $s$-process elements, particularly lead, instead of first or second $s$-process peak elements (like barium). However, lead is extremely hard to measure in carbon-enhanced metal-poor stars.

To summarise, there are ways for AGB stars to produce less barium and/or other s-process elements than usual. It is unclear what abundance patterns exactly are expected in extremely metal-poor AGB stars, and therefore what their companions that have received mass-transfer from such a star should look like. 

\subsection{A high fraction of binaries among intermediate/high carbon band CEMP-no stars}

Imagine a scenario in which a carbon-normal metal-poor star or an existing low-carbon band CEMP-no star in a binary system is (further) enriched in carbon by mass-transfer from such an AGB star, bringing them up to the intermediate/high A(C) band without enhancing their barium. Some or all of the carbon for these CEMP-no stars can be intrinsic, but also some or all of it can come from mass transfer from a former AGB companion. If a number of CEMP-no stars were additionally enhanced in carbon over their lifetime and some are only intrinsically carbon-enhanced, that is a possible explanation for the potential discrepancy in binary fraction between the higher A(C) and the low A(C) populations of CEMP-no stars. 

It is not likely that mass-transfer from a companion has happened in \textit{all} the intermediate/high A(C) CEMP-no binary stars. For example, SDSS~J0929$+$0238 with A(C) $= 7.44$ is a double-lined (possibly even triple-lined) spectroscopic binary of two main sequence stars with similar temperatures \citep{Caffau16}. If the carbon comes from mass-transfer from an AGB companion, these stars were born in a hierarchical triple system with the third star more massive and finally polluting the other two. This scenario decreases in likelihood if it would be confirmed that SDSS~J0929$+$0238 is a triple-lined system, which means it consists of three main sequence CEMP-no stars. In that case the system would have had to been born as a quadruple system with one star being more massive and having evolved through the AGB phase. Probably a better scenario is that these stars were born with an intrinsically high A(C), with the carbon formed in the previous generation of stars (as in the spinstar and/or faint supernova models). 

Interesting to note is that there are also five CEMP-$s$ stars that do not show any radial velocity variation. The binary fraction of CEMP-$s$ stars determined in \citet{Hansen16b}, $82 \pm 10 \%$, is not necessarily consistent with 100 \% binarity. They claim that even with the uncertainties on the inclination it is unlikely that all the apparently single stars are actually in binary systems. We note that in Figure~\ref{fig:AC} the single CEMP-$s$ stars seem to be preferentially located on the lower side of the A(C) distribution of the CEMP-$s$ stars. \citet{Choplin17} model the abundances of the single CEMP-s stars with massive spinstar models, and succeed for three out of the four modelled stars. Spinstar models are also employed to explain the abundances of CEMP-no stars (e.g. \citealt{Meynet10}). Potentially, single intermediate/high carbon band CEMP-no stars (or binaries that have not had any interaction) and single CEMP-s stars are the product of a similar type of progenitor.

A takeaway from this section is that a combination of one or more of the ``classical" scenarios for the formation of CEMP-no stars and binary interaction complicates the interpretation of the abundance patterns of these stars, usually thought to be direct probes of nucleosynthesis in the first stars and supernovae. 

\subsubsection{A scenario without binary interaction}

An alternative (or supplement) to the mass-transfer scenario is a scenario where binary stars form more easily in a carbon-enhanced environment. For example \citet{Chiaki17} have shown that for [C/Fe] $< +2.30$, silicate dust grains dominate the cooling during star formation of extremely metal-poor stars, while for [C/Fe] $> +2.30$ carbon grains dominate. It is not known how this would affect the binary fraction, but there is a possibility that there is a difference between these environments of different dust cooling. Unfortunately, little work has been done on the binary fraction of carbon-\textit{normal} extremely metal-poor stars and more observations are needed to study whether there is a difference in binary fraction between carbon-rich and carbon-normal extremely metal-poor environments.

\subsection{HE~0107$-$5240}

HE~0107$-$5240 is the most iron-deficient binary in our sample, and at the time of its discovery it was the most iron-deficient star known \citep{Christlieb02}. Since then, many different scenarios have been proposed to explain its chemical properties including its high carbon abundance. There are two main scenarios: 1) the abundance pattern of the star reflects the interstellar medium from which it was born which has been polluted by one or more primordial core-collapse supernovae, 2) the surface of the star has been polluted by material from a binary companion. So far none of the scenarios can completely explain the abundance pattern of HE~0107$-$5240, additionally it could also be a combination of the two.

In the first scenario there are different possible sources producing the necessary amount of carbon and the peculiar abundance pattern of HE~0107$-$5240, for example spinstars \citep{Takahashi14}, faint supernovae (\citealt{UmedaNomoto03}; \citealt{Iwamoto05}) or a combination of normal and faint supernovae \citep{Limongi03}. These scenarios seem to be relatively successful but their predictions are not entirely in agreement with the observations, especially for the oxygen abundance \citep{Bessell04}. 

The alternative scenario of a binary companion transfer has been investigated by \citet{Suda04}, \citet{Lau07} and \citet{Cruz13}. In the last work, the $s$-process abundance pattern and enhancement of carbon of HE~0107$-$5240 are explained by invoking mass transfer from a low-mass companion AGB star (both stars have some initial metallicity larger than zero). These models seem to fit the abundance pattern well, besides for nitrogen which is overproduced in the models. In the first two works, HE~0107$-$5240 starts out as a Population III (originally metal-free) star, where mass transfer from the companion star is fully responsible for the abundance pattern of HE~0107$-$5240. These models are also relatively successful at reproducing the chemical properties of this star. A prediction of the models is that the period of the binary is currently at least 30 years \citep{Lau07} and up to 150 years \citep{Suda04} with a maximum radial velocity variation of 6.5 -- 7~\kms. Their period range of $11000 - 55000$ days and semi-amplitude of $\sim 3.5$~\kms~are in good agreement with the results of this work, see the first panel of Figure~\ref{fig:orbits}. Additionally, \citet{Venn14} have found a marginal detection of mid-IR excess of this star. They speculate that if this excess is real, it might be a possible indication for a debris disk formed in a binary interaction.\footnote{Of the other stars in our sample, only SDSS~J0140$+$2344 seems to show a clear excess in the WISE Band 4. However, images reveal that 18~arcsec next to this star there is a bright QSO that completely dominates the WISE Band 4 image.}

In summary, our observations support the binary transfer model for the origin of HE~0107$-$5240, where this star is potentially a true first generation star whose pristine atmosphere has been spoiled during its lifetime. In the binary transfer scenario it can also be a second generation star whose surface is additionally polluted by mass-transfer from a companion, which would complicate the interpretation of its abundance pattern.

\begin{figure}
\centering
\includegraphics[width=\hsize,trim={0.1cm 0.4cm 0.5cm 0.0cm}]{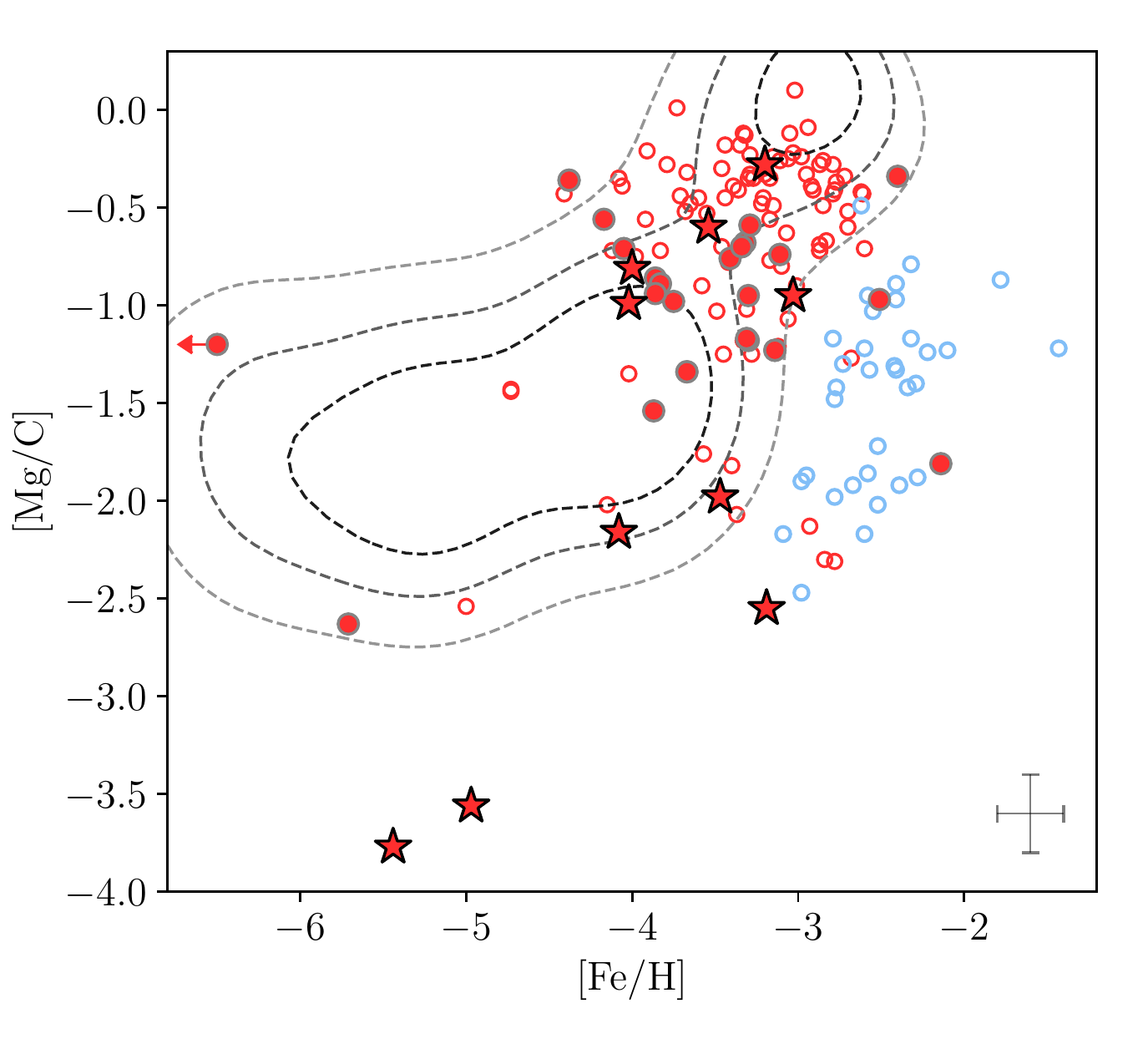}
\caption{Following \citet{Hartwig18}, we present [Mg/C] values for CEMP-no and CEMP-$s$ stars against [Fe/H]. The symbols are the same as in Figure~\ref{fig:AC} and \ref{fig:HR}, while open symbols represent stars without binarity information. Typical error bars are indicated in the lower right corner. Contours of likely mono-enrichment are over-plotted (priv. comm. T. Hartwig).}
 \label{fig:MgC}
\end{figure}

\subsection{Magnesium}

In their comparison of the two different CEMP-no sub-groups, \citet{Yoon16} look at magnesium and find that in A(C) vs. A(Mg) space the Group II stars scale roughly with A(C), whereas the Group III stars do not and appear offset to lower A(Mg) values. Based on this behaviour (and similar behaviours for other elements) they suggest that the Group II and Group III CEMP-no stars could be associated to different classes of progenitors, possibly the faint mixing-and-fallback supernovae and spinstar respectively. They do mention that this is a tentative conclusion and that there may be other factors at play. Several of the CEMP-no binaries fall in the Group III subclass, suggesting that binarity may be one of those other factors. 

Recently, \citet{Hartwig18} have presented a novel diagnostic to identify second generation stars whose birth gas cloud was enriched by only one supernova. They use the so-called ``divergence of the chemical displacement" to identify regions in chemical space where it is likely to find these kind of stars. This divergence does not depend on many assumptions besides the input supernovae yields for core-collapse, pair-instability and faint supernovae. They do not have rotating first stars or any mass-transfer scenarios as possible sources of metals in their model, but mention that they plan to add these in later work. One of the abundance spaces they identify as useful is [Fe/H] vs. [Mg/C]. 

To see what the CEMP stars look like in this space, we compile their magnesium abundances. For the single and binary CEMP-no stars, [Mg/C] values are listed in Table~\ref{table:rvcemplist}. For further Group II and III stars we use the compilation of \citet{Yoon16}, which we supplemented with [Mg/Fe] for Group I stars. For the CEMP-$s$ stars, we compiled [Mg/Fe] from \citet{Yong13} and \citet{Roederer14} to get a representative sample of CEMP-$s$ stars for comparison (both use the \citet{Asplund09} solar abundances). We cross-matched those stars with \citet{Yoon16} to get the corrected carbon abundances for consistency. For the CEMP-$s$ stars, we do not include the binary information since there are only a few stars with binarity information in the combined Yong and Roederer sample, and it can be assumed that almost all of them are in fact in binary systems. 

The result is shown in Figure~\ref{fig:MgC}. The CEMP-$s$ stars all lie outside the range where \citet{Hartwig18} claim mono-enrichment is likely. This is not surprising since they are expected to have received mass-transfer from a binary companion, and the models do not include this. The CEMP-$s$ stars are offset towards lower [Mg/C] compared to most of the CEMP-no stars at similar [Fe/H], which is consistent with their formation scenario of binary transfer from a former AGB companion, which mainly enhances C in the companion star and not Mg. This diagram may be a useful tool in uncovering stars that have undergone mass-transfer. 

Most of the CEMP-no stars lie in the mono-enrichment range. Five of the CEMP-no binary stars however have relative low values of [Mg/C] ($\lesssim-2.0$). Like the CEMP-$s$ stars, these stars may have experienced mass transfer from a binary companion which enhanced C but not Mg. These five stars are HE~0107$-$5240, SDSS~J0929$+$0238, G77$-$61, HE~1150$-$0428 and CS~22957$-$027 (in order of increasing [Fe/H]). For HE~0107$-$5240, the very low [Mg/C] ratio was also reproduced in the mass-transfer model by \citet{Cruz13}. G77$-$61 most likely has a white dwarf as a companion \citep{Dearborn86}, so past mass transfer is also not unlikely. For SDSS~J0929$+$0238 we argued earlier that the mass-transfer scenario is unlikely (see Section~\ref{sec:bintransfer}), however it is still possible. If it is has not experienced AGB mass transfer, some other explanation needs to be found for its low [Mg/C]. 

Worth noting is that all CEMP-no stars that overlap with the location of the CEMP-$s$ stars in this diagram are Group~I stars, regardless their binary status. CEMP-no stars in Group~I have A(C) and [Fe/H] similar to those of the CEMP-$s$ stars, but low [Ba/Fe]. Where these stars get such high carbon abundances from is unclear, especially for the single stars, but it is interesting that also in this space they share properties with CEMP-$s$ stars. However, it also shows that interpretation of this diagram is not trivial and we should be careful to draw strong conclusions.

Finally, we emphasise that further caution should be taken when interpreting this figure, since most of the abundance measurements were not computed using non-LTE and/or 3D models and such corrections can be important both for C and Mg.

\section{Radial velocity outlook with\textit{ Gaia}}
\label{sec:Gaia}

\begin{table*}
\caption{\label{table:Gaiastars} CEMP stars that have Gaia radial velocity uncertainties $> 1$ \kms, with new binary candidates in bold-face.}
\begin{tabular}{llrrrrrrrrrrr}
 \hline
Name  &  CEMP  &  G  &  $rv_\mathrm{Gaia}$  &  $\sigma_\mathrm{Gaia}$  &  $rv_\mathrm{lit}$\tablefootmark{a}   &  $\sigma_\mathrm{lit}$  &  n$_\mathrm{lit}$  &  ref &  $T_\mathrm{eff}$\tablefootmark{b} &  [Fe/H]\tablefootmark{b}  & $T_\mathrm{eff,T}$  & [Fe/H]$_\mathrm{T}$ \\
  &  class  &  \tiny{(mag)}  &  \tiny{(km$\,$s$^{-1}$)}  &  \tiny{(km$\,$s$^{-1}$)} &  \tiny{(km$\,$s$^{-1}$)} &  \tiny{(km$\,$s$^{-1}$)}  &    &    &  \tiny{(K)}  &    & \tiny{(K)}    &  \\
 \hline
\textbf{HE 2155$-$3750}  &  s  &  13.1  &  77.5  &  4.5  &  27.7  &  10.0\tablefootmark{c}  &  1  &  1  &  5000  &  $-$2.64  &  5500  &  0.0  \\
\textbf{HE 1305$+$0007}  &  s  &  11.9  &  228.5  &  3.3  &  217.8  &  1.5  &  1  &  2  &  4750  &  $-$2.08  &  4750  &  $-$2.0  \\
HE 2123$-$0329  &  no  &  13.1  &  $-$217.7  &  3.1  &  $-$219.1  &  0.4  &  2  &  3  &  4725  &  $-$3.22  &  6500  &  $-$1.5  \\
\textbf{SMSS J1738$-$1457}   &  no  &  12.9  &  43.7  &  3.1  &  $-$26.9  &  2.0  &  1  &  4  &  4600  &  $-$3.58  &  5700  &  0.0  \\
HKII 17435$-$00532  &  r/s  &  12.8  &  39.3  &  2.4  &  38.9  &  0.3  &  4  &  5  &  5200  &  $-$2.23  &  6000  &  $-$1.5  \\
CS 30301$-$015   &  s  &  12.8  &  87.0  &  1.8  &  86.6  &  0.1  &  18  &  6  &  4900  &  $-$2.73  &  6000  &  $-$1.5  \\
\textbf{HE 2319$-$5228}  &  no  &  13.0  &  286.2  &  1.8  &  294.3  &  4.0  &  2  &  7  &  4900  &  $-$2.60  &  6000  &  $-$1.5  \\
BS 16077$-$077  &  s  &  11.9  &  68.3  &  1.7  &  ...  &  ...  &  ...  &  ...  &  5900  &  $-$2.05  &  6000  &  $-$1.5  \\
CS 22947$-$187  &  s  &  12.8  &  $-$252.3  &  1.4  &  $-$251.9  &  0.9  &  2  &  8,9  &  5300  &  $-$2.58  &  6000  &  $-$1.5  \\
HE 1305$+$0132  &  s  &  12.4  &  157.0  &  1.4  &  ...  &  ...  &  ...  &  ...  &  4462  &  $-$2.45  &  4500  &  0.0  \\
CS 22873$-$128  &  no  &  12.8  &  207.15  &  1.2  &  205.5  &  0.3  &  2  &  8,9  &  4710  &  $-$3.32  &  6000  &  $-$1.5  \\

 \hline
\end{tabular}

\tablefoot{References: (radial velocities) (1) \citet{Placco11}, (2) \citet{Goswami06}, (3) \citet{Hollek11}, (4) \citet{Jacobson15}, (5) \citet{Roederer08}, (6) \citet{Hansen16b}, (7) \citet{Beers17}, (8) \citet{Roederer14}, (9) \citet{McWilliam95} \\
\tablefoottext{a}{For stars with only one measurement, this is the radial velocity as presented in the literature. For stars with multiple measurements (see column n$_\mathrm{lit}$) it is the (weighted) average of the different measurements. Similarly for the uncertainty, it is either the uncertainty reported in the literature or the (weighted) standard deviation between different literature values. } \\
\tablefoottext{b}{As reported in the \citet{Yoon16} compilation.} \\
\tablefoottext{c}{No uncertainty is given in the reference, but since it is a measurement from an intermediate resolution spectrum we assume $\sigma$ = 10 \kms. }
}

\end{table*}

More data are needed to increase the sample of CEMP-no stars with multiple radial velocity measurements and put more stringent constraints on the orbits for several of the CEMP-no binaries. This might also shed light on the evolutionary status of the companion stars and constrain mass transfer models. Additionally, it can be the case that there are more long-period variable stars hiding in the current sample. The ESA \textit{Gaia} mission will have several epochs of radial velocity data for all the brightest stars in the Galaxy down to $ V \approx 16.2$ \citep{Gaia16}. For the faintest metal-poor ([Fe/H] $ = -1.5$) stars, \textit{Gaia} will have end-of-mission radial velocity uncertainties larger than 15~\kms, whereas for stars brighter than $V = 14$ the expected uncertainty is between $0.5 - 2$~\kms.\footnote{\url{https://www.cosmos.esa.int/web/gaia/rvsperformance}} This is unfortunately not the uncertainty on the individual radial velocity measurements, but that of the combined end-of-mission radial velocity. Additionally, only for the brightest stars the single epoch radial velocities will probably be released in a future \textit{Gaia} data release. It is unclear how much \textit{Gaia} will contribute to providing multiple good radial velocity measurements that can constrain orbits of CEMP stars, therefore it is still important to continue the radial velocity monitoring effort with high resolution spectrographs here on Earth.

However, one can use \textit{Gaia} data to find new binary systems, even in \textit{Gaia} DR2 \citep{Gaia18}. The radial velocity uncertainties provided in DR2 are the result of the combination of multiple radial velocity measurements, and if stars vary in radial velocity over the course of the \textit{Gaia} observations they will have higher radial velocity uncertainties than expected from the precision for stars of their magnitude and effective temperature. This approach is similar to what was done to investigate binarity in CEMP stars discovered in the APOGEE survey \citep{Kielty17}.

\begin{figure}
\centering
\includegraphics[width=\hsize,trim={0.5cm 0.4cm 0.5cm 0.0cm}]{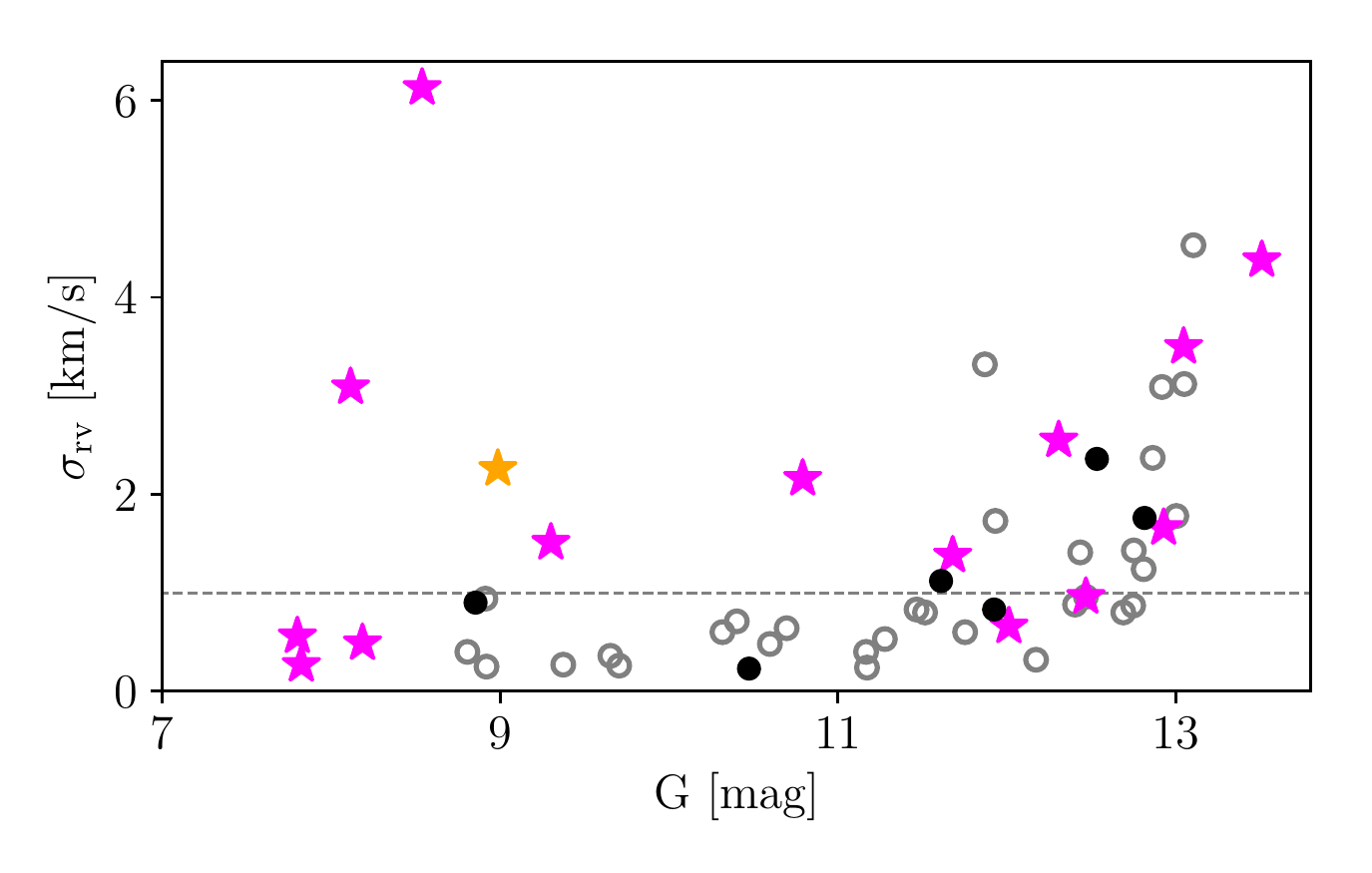}
\caption{Radial velocity uncertainty from \textit{Gaia} versus the \textit{Gaia G} magnitude. Stars are labelled by binarity using the information from the \citet{Yoon16} compilation, where magenta stars are binaries, black circles are non-binaries and open grey circles represent stars of unknown binarity. The orange star is HD~135148. At fainter magnitudes, the radial velocity uncertainties in \textit{Gaia} increase. The 1 \kms~limit is marked by the grey dotted line.}
 \label{fig:Gaia}
\end{figure}

An example of this type of analysis is shown in Figure~\ref{fig:Gaia}, where we cross-matched the \citet{Yoon16} CEMP star sample with \textit{Gaia} DR2 and present the uncertainty in radial velocity versus the $G$ magnitude for the stars sufficiently bright to be included in the current data release. At the faintest end, the precision is expected to be a maximum of $\sim 2$~\kms~for the hottest stars in our sample, and at the bright end the precision should be less than 1~\kms~\citep{Katz18}. In the figure we indicate binarity information from the compilation, where the magenta star symbols are known binaries, the black circles are known non-binaries and open grey symbols represent stars of unknown binarity. The orange star in the Figure is HD~135148, for which the binary information was not provided in the \citet{Yoon16} compilation. We ``rediscovered" this binary using the \textit{Gaia} information and found that it was already known from \citet{Carney03}. It is clear that several of the binaries indeed have larger radial velocity uncertainties than expected from the precision alone. 

\subsection{New CEMP binary candidates}

We investigate stars of unknown binarity that have large $\sigma_\mathrm{Gaia}$ values ($ > 1$~\kms), see a summary in Table~\ref{table:Gaiastars}. We included the $T_\mathrm{eff,T}$ and [Fe/H]$_\mathrm{T}$ of the template that was used in the Gaia radial velocity determination. A bad template with the wrong shape of the spectral lines might not necessarily result in a bad radial velocity, but it will most probably influence the radial velocity uncertainty. The standard Gaia radial velocity templates have [Fe/H] of either 0.0 or $-$1.5, except when the metallicity of the star is known \citep{Katz18}. Even though most of our stars are more metal-poor than [Fe/H] $= -1.5$, the radial velocity precision seems to be good with the [Fe/H]~$= -1.5$ templates, since several stars in Table~\ref{table:Gaiastars} have literature and \textit{Gaia} radial velocities agreeing to within 1~\kms. This is even the case for stars where $T_\mathrm{eff,T}$ is more than 1000~K off from the literature (see CS~30301$-$015 and CS~22873$-$128). It is unclear how good the velocities are with the [Fe/H]~=~0.0 templates, since two stars with such a template have literature radial velocities highly discrepant with those from \textit{Gaia}. In future \textit{Gaia} data releases the stellar parameters from the \textit{Gaia} spectra (Bp, Rp and RVS) will be used to select better radial velocity templates, which should reduce the mismatch between observations and templates and improve the radial velocity quality \citep{Katz18}. 

In the cross-match between the \citet{Yoon16} CEMP sample and the \textit{Gaia} data there are two CEMP-$s$ stars without currently available binarity information, HE~2155$-$3750 and HE~1305$+$0007, that have large $\sigma_\mathrm{Gaia}$ and large discrepancies (>10~\kms) between the \textit{Gaia} radial velocity and the literature. It should be noted that the [Fe/H]$_\mathrm{T}$ of HE~2155$-$3750 is a bad match, which may cause part of the discrepancy but unlikely the full $\sim$50~\kms. These two stars seem good binary~candidates.

Additionally, and of more interest for this work, there are two CEMP-no stars that seem to vary in radial velocity, SMSS~J1738$-$1457 and HE~2319$-$5228. In the case of SMSS~J1738$-$1457, the applied radial velocity template is a bad match in both [Fe/H] and $T_\mathrm{eff}$. It is however unlikely that this causes a difference of $\sim 60$~\kms, therefore this star is still a good binary candidate. With its A(C)~=~6.18 and [Fe/H]~=~$-3.58$ \citep{Jacobson15}, it lies on the lower carbon band. Then HE~2319$-$5228 is showing a modest radial velocity variation of $\sim 8$~\kms. We have seen that a mismatch between $T_\mathrm{eff}$ and $T_\mathrm{eff,T}$ is not likely causing such large differences (e.g. compare to CS~30301$-$015), therefore this star remains a good binary candidate. It has A(C)~=~6.51 and [Fe/H]~=~$-3.4$ \citep{Beers17}, therefore it also lies on the lower carbon band. 

Another way to find binary stars using \textit{Gaia} is using the astrometric excess noise (D) and the goodness of fit of the \textit{Gaia} astrometry, to find stars with bad astrometric solutions caused by a companion star (as demonstrated e.g. in \citealt{Evans18}). This method works best for nearby stars, and unfortunately most of the extremely metal-poor CEMP-no stars are too far away. However, there are two CEMP-no stars in the \citet{Yoon16} sample, G77$-$61 and CS~22958$-$042, that have significant astrometric excess noise (D~$ > 3$). G77$-$61 is a nearby, high proper motion, known binary star which has a bad goodness of fit and D~$=~207.9$. CS~22958$-$042 has D~$=~9.7$ and also a bad goodness of fit, therefore this star might well be in a binary system too.

The analysis in this section shows that \textit{Gaia} can be used to find new candidate binary CEMP systems, however follow-up spectroscopy is still needed to verify the results and characterise the orbits of the new binary systems. 

\section{Conclusions}
\label{sec:conclusions}

In this paper we have discussed the results of the extension of a radial velocity monitoring program for CEMP-no stars that started with \citet{Starkenburg14}. We have identified four new CEMP-no stars in binary systems based on their radial velocity variations. Together with stars from the literature we now have a sample of eleven CEMP-no binaries and twenty-three likely single CEMP-no stars, resulting in a binary fraction of $32^{+10\,}_{\,\,-9}\%$. This only marginally agrees with the previous estimate for CEMP-no stars by \citet{Hansen16a}, which was similar to the binary fraction of metal-poor carbon-normal giants ($16^{+5\,}_{-4} \%$, \citealt{Carney03}). 

The periods of the CEMP-no binaries are similar to the typical periods of CEMP-$s$ stars, ranging from several 100 to several 1000 days for most of the stars. It applies to all the binaries in our sample that if the companion of the main star is currently a white dwarf, it is probable that the current CEMP-no star has been polluted during the AGB phase of the companion. This enhances the amount of carbon in the star and changes its abundance pattern. None of the CEMP-no binary stars show a clear enhancement in [Ba/Fe] indicative of s-process element transfer. Therefore, if binary transfer from an AGB companion has happened, it must have been an extremely metal-poor AGB star that has not produced a significant amount of s-process elements.

Within our small CEMP-no radial velocity sample there is an apparent difference in binary fraction between the CEMP-no stars with lower A(C)~($18^{\,+14\,}_{\,\,-9} \%$) and those with higher A(C)~($47^{\,+15\,}_{\,-14}\%$). This higher binary fraction of stars with high absolute carbon abundances can have interesting implications for the origins of CEMP-no stars. We propose that some of the high A(C) binaries started out as carbon-normal stars or CEMP-no stars on the low carbon band and received extra carbon from a companion moving them to the intermediate/high carbon band. Alternatively, star formation might have a tendency to form more binaries at high A(C) or extremely low [Fe/H]. 

Especially interesting is the detection of radial velocity variations in the hyper metal-poor star HE~0107$-$5240. Some models have tried to explain its abundance pattern based on the assumption that its completely pristine surface has been polluted by a former AGB companion, currently a white dwarf. Previous to this current detection, there was no clear evidence for its variability in radial velocity. This makes the binary formation scenario as probable as, or even more probable than, scenarios invoking faint supernovae or spinstars. 

Further monitoring of our new CEMP-no binary stars would allow the derivation of better orbit parameters, which is necessary to constrain possible mass transfer models. Monitoring of additional stars is needed to investigate wether there truly is a larger fraction of CEMP-no binary stars among stars with high absolute carbon abundance and/or extremely low metallicity. Although \textit{Gaia} may not (yet) be instrumental in studying specific binary systems in detail or in constraining the binary fraction, it shows promise in discovering new binary systems. We have highlighted some new binary candidates.

Some (although not necessarily all) CEMP-no binary stars might have been polluted by a companion star, which affects and complicates the interpretation of their abundance patterns. Their abundances may not only be probes of faint supernovae and/or spinstars but also of extremely metal-poor AGB stars. 

\section*{Acknowledgements}

AA and ES gratefully acknowledge funding by the Emmy Noether program from the Deutsche Forschungsgemeinschaft (DFG). KAV thanks the National Science and Engineering Research Council for partial funding through their Discovery Grants program. We thank Lison Malo for her help with the CFHT ESPaDOnS spectra, Norbert Christlieb for sharing his radial velocity data for HE~0557$-$4840 and HE~0107$-$5240 with us, Terese Hansen for her re-determination of the stellar parameters for HE~0219$-$1739 and Tilman Hartwig for sharing his mono-enrichment contours for Figure~\ref{fig:MgC}. We thank John Norris, Jay Farihi and Terese Hansen for their useful comments on a draft of this work. The observations reported in this paper were obtained with the Southern African Large Telescope (SALT) and the Canada France Hawaii Telescope (CFHT). It also made use of data obtained from the ESO Science Archive Facility under request numbers \textit{aarentsen} 312315 and 358299. This work has made use of data from the European Space Agency (ESA) mission {\it Gaia} (\url{https://www.cosmos.esa.int/gaia}), processed by the {\it Gaia} Data Processing and Analysis Consortium (DPAC,
\url{https://www.cosmos.esa.int/web/gaia/dpac/consortium}). Funding for the DPAC has been provided by national institutions, in particular the institutions participating in the {\it Gaia} Multilateral Agreement. This research made extensive use of the \textsc{matplotlib} \citep{Hunter07} and \textsc{pandas} \citep{McKinney10} Python packages.


\bibliographystyle{aa}
\bibliography{CEMPpaper.bbl}   

\begin{appendix}

\section{Measured radial velocities}

\begin{table}[H]
\caption{\label{table:rvs} Radial velocities determined from the CFHT and SALT spectra in this work.}
\begin{tabular}{lrrrl}
 \hline
 Name & rv & err & HJD & flag\tablefootmark{a}\\
   & (km$\,$s$^{-1}$) & (km$\,$s$^{-1}$) & $-2450000$  &  \\
 \hline
BD$+$44$^{\circ}$493 & $-$149.77 & 0.55 & 6525.58 & C \\
BD$+$44$^{\circ}$493 & $-$150.24 & 0.51 & 6549.55 & C \\
BD$+$44$^{\circ}$493 & $-$149.95 & 0.35 & 6697.25 & C \\
BD$+$44$^{\circ}$493 & $-$149.32 & 1.32 & 6709.21 & C \\
BD$+$44$^{\circ}$493 & $-$149.62 & 1.13 & 6709.21 & C \\
 & & & & \\
BS 16929$-$005 & $-$50.84 & 0.54 & 6523.22 & C \\
BS 16929$-$005 & $-$50.56 & 0.76 & 6675.68 & C \\
BS 16929$-$005 & $-$51.27 & 0.90 & 6701.62 & C \\
BS 16929$-$005 & $-$49.70 & 0.60 & 6701.63 & C \\
 & & & & \\
CS 22878$-$027 & $-$91.77 & 0.35 & 6520.32 & C \\
CS 22878$-$027 & $-$91.45 & 0.34 & 6530.23 & C \\
CS 22878$-$027 & $-$91.56 & 0.51 & 6700.64 & C \\
CS 22878$-$027 & $-$91.62 & 0.34 & 6812.32 & C \\
 & & & & \\
CS 22949$-$037 & $-$126.10 & 0.44 & 6521.53 & C \\
CS 22949$-$037 & $-$126.12 & 0.41 & 6532.54 & C \\
CS 22949$-$037 & $-$126.05 & 0.56 & 6549.50 & C \\
CS 22949$-$037 & $-$125.67 & 0.32 & 6816.59 & C \\
CS 22949$-$037 & $-$126.10 & 0.36 & 6824.54 & C \\
 & & & & \\
CS 22957$-$027 & $-$61.43 & 0.94 & 6519.53 & C \\
CS 22957$-$027 & $-$62.34 & 0.89 & 6532.55 & C \\
CS 22957$-$027 & $-$61.46 & 0.92 & 6549.52 & C \\
CS 22957$-$027 & $-$71.68 & 1.05 & 6816.61 & C \\
CS 22957$-$027 & $-$72.09 & 0.98 & 6824.56 & C \\
 & & & & \\
CS 29498$-$043 & $-$30.04 & 3.88 & 6609.25 & S \\
CS 29498$-$043 & $-$31.78 & 1.32 & 6876.25 & S \\
CS 29498$-$043 & $-$31.61 & 2.70 & 6968.25 & S \\
CS 29498$-$043 & $-$32.24 & 0.36 & 7121.64 & S \\
 & & & & \\
CS 29502$-$092 & $-$67.33 & 0.35 & 6519.53 & C \\
CS 29502$-$092 & $-$66.91 & 0.38 & 6530.43 & C \\
CS 29502$-$092 & $-$67.55 & 0.33 & 6548.33 & C \\
CS 29502$-$092 & $-$66.93 & 0.31 & 6770.64 & C \\
CS 29502$-$092 & $-$67.54 & 0.35 & 6770.64 & C \\
CS 29502$-$092 & $-$66.54 & 0.26 & 6816.58 & C \\
 & & & & \\
 HE 0057$-$5959 & 377.90 & 3.98 & 6611.50 & S \\
HE 0057$-$5959 & 378.23 & 1.47 & 6886.50 & S \\
 & & & & \\
HE 0107$-$5240 & 48.15 & 0.49 & 6618.50 & S \\
HE 0107$-$5240 & 46.60 & 4.75 & 6858.50 & S \\
HE 0107$-$5240 & 47.19 & 2.39 & 6957.25 & S \\
HE 0107$-$5240 & 48.29 & 0.94 & 6990.50 & S \\
 & & & & \\
HE 0557$-$4840 & 209.11 & 2.00 & 6644.25 & S \\
HE 0557$-$4840 & 212.33 & 0.58 & 6923.50 & S \\
 & & & & \\
 HE 1012$-$1540 & 226.17 & 0.64 & 6669.61 & C \\
HE 1012$-$1540 & 225.33 & 0.59 & 6669.62 & C \\
HE 1012$-$1540 & 225.27 & 0.34 & 6675.60 & C \\
HE 1012$-$1540 & 225.35 & 0.33 & 6697.51 & C \\
HE 1012$-$1540 & 225.41 & 0.26 & 6758.22 & C \\
HE 1012$-$1540 & 225.43 & 0.49 & 6998.50 & S \\
  & & & & \\
\hline
\end{tabular}
\end{table}

\setcounter{table}{0}

\begin{table}[H]
\caption{\label{table:rvs2} (continued)}
\begin{tabular}{lrrrl}
 \hline
 Name & rv & err & HJD & flag\tablefootmark{a} \\
   & (km$\,$s$^{-1}$) & (km$\,$s$^{-1}$) & $-2450000$ & \\
 \hline

HE 1150$-$0428 & 38.17 & 1.70 & 6669.64 & C \\
HE 1150$-$0428 & 36.31 & 1.38 & 6675.62 & C \\
HE 1150$-$0428 & 35.90 & 1.59 & 6697.53 & C \\
HE 1150$-$0428 & 46.23 & 1.51 & 6770.44 & C \\
HE 1150$-$0428 & 54.06 & 1.55 & 6812.25 & C \\
HE 1150$-$0428 & 54.49 & 1.62 & 6817.23 & C \\
HE 1150$-$0428 & 55.91 & 1.65 & 6823.24 & C \\
 & & & & \\
HE 1201$-$1512 & 239.80 & 1.32 & 6669.67 & C \\
HE 1201$-$1512 & 237.06 & 1.70 & 6669.67 & C \\
HE 1201$-$1512 & 240.11 & 0.52 & 6697.56 & C \\
HE 1201$-$1512 & 240.07 & 0.79 & 6760.26 & C \\
HE 1201$-$1512 & 235.18 & 3.34 & 7058.47 & S \\
HE 1201$-$1512 & 239.26 & 6.04 & 7141.24 & S \\
 & & & & \\
HE 1300+0157 & 75.40 & 1.06 & 6520.22 & C \\
HE 1300+0157 & 72.89 & 1.20 & 6675.67 & C \\
HE 1300+0157 & 74.60 & 0.44 & 6700.53 & C \\
HE 1300+0157 & 74.67 & 0.72 & 6760.28 & C \\
 & & & & \\
HE 1327$-$2326 & 65.35 & 1.60 & 6781.25 & S \\
HE 1327$-$2326 & 64.15 & 1.35 & 7060.50 & S \\
HE 1327$-$2326 & 67.86 & 3.52 & 7140.53 & S \\
 & & & & \\
 HE 1506$-$0113 & $-$93.91 & 0.54 & 6520.24 & C \\
HE 1506$-$0113 & $-$92.80 & 0.46 & 6527.22 & C \\
HE 1506$-$0113 & $-$92.75 & 0.51 & 6532.22 & C \\
HE 1506$-$0113 & $-$87.43 & 0.46 & 6697.63 & C \\
HE 1506$-$0113 & $-$88.23 & 0.46 & 6701.64 & C \\
HE 1506$-$0113 & $-$85.32 & 0.34 & 6812.29 & C \\
HE 1506$-$0113 & $-$84.61 & 0.37 & 6823.27 & C \\
HE 1506$-$0113 & $-$77.13 & 1.97 & 7140.54 & S \\
 & & & & \\
HE 2139$-$5432 & 98.58 & 4.00 & 6620.25 & S \\
HE 2139$-$5432 & 95.79 & 0.29 & 6900.25 & S \\
 & & & & \\
HE 2142$-$5656 & 104.24 & 0.82 & 6609.25 & S \\
HE 2142$-$5656 & 104.06 & 1.98 & 6876.25 & S \\
HE 2142$-$5656 & 105.32 & 2.25 & 6968.25 & S \\
 & & & & \\
HE 2202$-$4831 & 56.58 & 1.20 & 6609.25 & S \\
HE 2202$-$4831 & 57.50 & 2.00 & 6990.25 & S \\
 & & & & \\
HE 2247$-$7400 & 6.54 & 1.53 & 6881.50 & S \\
 & & & & \\
SDSS J0140+2344 & $-$190.32 & 1.09 & 6525.55 & C \\
SDSS J0140+2344 & $-$201.03 & 1.25 & 6549.52 & C \\
SDSS J0140+2344 & $-$187.99 & 1.69 & 6561.38 & C \\
SDSS J0140+2344 & $-$202.03 & 0.82 & 6824.57 & C \\
 & & & & \\
SDSS J1422+0031 & $-$125.43 & 2.26 & 6717.50 & S \\
SDSS J1422+0031 & $-$113.96 & 2.32 & 7071.59 & S \\
SDSS J1422+0031 & $-$112.35 & 0.23 & 7122.52 & S \\
\hline
\end{tabular}

\tablefoot{\tablefoottext{a}{C is observed with CFHT, S is observed with SALT}}

\end{table}

\newpage

\section{Compilation of 45 CEMP stars with multiple radial velocity measurements}

\begin{table*}[h!]
\caption{\label{table:rvcemplist} List of 45 CEMP-no stars with multiple radial velocity measurements.}
\resizebox{\textwidth}{!}{%
\begin{tabular}{lrrrrlrrrrrrrrr}
\hline
 Name & n$_\mathrm{rv}$ & rv$_\mathrm{w}$  & $\sigma_\mathrm{rv}$ & $P(\chi^2)$ & bin? & $T_\mathrm{eff}$ & $\log g$ & [Fe/H] & [C/Fe] & A(C) & [Ba/Fe] & [Mg/C] & ref \\
  &  & (km$\,$s$^{-1}$) & (km$\,$s$^{-1}$) & & & (K) & & & & & & \\
\hline
BD$+$44$^{\circ}$493   &61& $-$150.09 &0.63&0.015&1&5430&3.4& $-$3.83 &1.35&5.95& $-$0.60 & $-$0.89 & 1\\
BS 16929$-$005   &21& $-$50.77 &0.66&0.629&1&5229&2.61& $-$3.34 &0.99&6.09& $-$0.41 & $-$0.70 & 2\\
CD$-$24$^{\circ}$17504   &2&136.25&0.49&0.227&1&6228&3.9& $-$3.41 &1.1&6.12& <$-$1.05 & $-$0.76 & 3\\
CS 22166$-$016   &9& $-$210.39 &0.77&0.607&1&5250&2& $-$2.40 & 1.02\tablefootmark{a} & 7.05\tablefootmark{a} & $-$0.37 & $-$0.34\tablefootmark{a} & 4\\
CS 22877$-$001   &16&166.25&0.12&0.790&1&4790&1.45& $-$3.31 &1.1&6.67& $-$0.50 & $-$1.17 & 5\\
CS 22878$-$027   &21& $-$91.28 &0.61&0.058&1&6319&4.41& $-$2.51 & 0.86\tablefootmark{a} & 6.78\tablefootmark{a} & <$-$0.75 & $-$0.97\tablefootmark{a} & 2\\
CS 22949$-$037   &26& $-$125.74 &0.29&0.889&1&4630&0.95& $-$4.38 &1.16&5.97& $-$0.60 & $-$0.36 & 5\\
CS 22957$-$027   &40& $-$66.87 &5.16&$<10^{-6}$&2&5220&2.65& $-$3.19 &2.61&7.87& $-$0.81 & $-$2.55 & 5\\
CS 29498$-$043   &26& $-$32.55 &0.6&0.994&1&4440&0.5& $-$3.87 &2.75&7.62& $-$0.49 & $-$1.54 & 5\\
CS 29502$-$092   &35& $-$67.22 &0.51&0.575&1&4820&1.5& $-$3.30 &1.06&6.59& $-$1.36 & $-$1.18 & 5\\
CS 29527$-$015   &6&47.13&0.42&0.988&1&6577&3.89& $-$3.32 &1.18&6.29&  & $-$0.68 & 2\\
G64$-$12   &33&442.55&1.05&0.009&1&6463&4.26& $-$3.29 & 1.07\tablefootmark{a} & 6.21\tablefootmark{a} & $-$0.07 & $-$0.59\tablefootmark{a} & 6\\
G64$-$37   &22&81.49&0.95&0.312&1&6570&4.4& $-$3.11 & 1.12\tablefootmark{a} & 6.44\tablefootmark{a} & $-$0.36 & $-$0.74\tablefootmark{a} & 6\\
G77$-$61   &13& $-$23.94 &11.35&$<10^{-6}$&2&4000&5.05& $-$4.08 &2.65&7& <$+$1.00 & $-$2.16 & 7\\
HE 0020$-$1741   &10&93.06&0.83&0.001&1&4765&1.55& $-$4.05 &1.4&6.12& $-$1.11 & $-$0.71 & 6\\
HE 0057$-$5959   &3&375.64&1.6&0.143&  &5257&2.65& $-$4.08 &0.86&5.21& $-$0.46 & $-$0.35 & 2\\
HE 0107$-$5240   &7&44.78&1.91&$<10^{-6}$&2&5100&2.2& $-$5.44 &3.97&7.03& <$+$0.93 & $-$3.77 & 8\\
HE 0219$-$1739   &15&107.96&5.09&$<10^{-6}$&2&4238&0.47& $-$3.09 &1.9&7.55& <$-$1.39 &  & 9\\
HE 0405$-$0526   &13&165.66&0.04&1.000&1&5083&3.86& $-$2.18 &0.92&7.17& $-$0.22 &  & 9\\
HE 0557$-$4840   &22&211.94&0.8&0.965&  &4900&2.2& $-$4.73 &1.59&5.29& <$+$0.07 & $-$1.43 & 10\\
HE 1012$-$1540   &20&225.84&0.47&0.217&1&5230&2.65& $-$4.17 &2.4&6.67& $-$0.28 & $-$0.56 & 5\\
HE 1133$-$0555   &9&270.7&0.34&0.810&1&5526&1.31& $-$2.40 &2.2&8.31& $-$0.58 &  & 9\\
HE 1150$-$0428   &27&47.49&8.13&$<10^{-6}$&2&5208&2.54& $-$3.47 &2.37&7.35& $-$0.48 & $-$1.98 & 2\\
HE 1201$-$1512\tablefootmark{b} &13&238.91&2.51&0.016&1&5725&4.67& $-$3.86 &1.14&5.71& <$+$0.05 & $-$0.86 & 2\\
HE 1300$-$0641   &2&68.79&0.11&0.162&1&5308&2.96& $-$3.14 &1.25&6.54& $-$0.82 & $-$1.23 & 11\\
HE 1300$+$0157   &20&74.4&0.7&0.871&1&5529&3.25& $-$3.75 &1.31&5.99& <$-$0.85 & $-$0.98 & 2\\
HE 1302$-$0954   &3&32.56&0.04&0.973&1&5120&2.4& $-$2.25 &1.17&7.37& <$-$0.53 &  & 9\\
HE 1327$-$2326   &17&64.38&1.3&0.202&1&6180&3.7& $-$5.71 &4.18&6.9& <$+$1.39 & $-$2.63 & 12\\
HE 1410$+$0213   &23&81.12&0.18&0.614&1&5000&2& $-$2.14 &1.92&8.21& $-$0.26 & $-$1.81 & 13\\
HE 1506$-$0113   &29& $-$83.50 &11.22&$<10^{-6}$&2&5016&2.01& $-$3.54 &1.47&6.38& $-$0.80 & $-$0.60 & 2\\
HE 2139$-$5432   &4&102.18&10.07&$<10^{-6}$&2&5416&3.04& $-$4.02 &2.59&7.01& <$-$0.33 & $-$0.99 & 2\\
HE 2142$-$5656   &4&103.7&0.8&0.718&  &4939&1.85& $-$2.87 &0.95&6.61& $-$0.63 & $-$0.72 & 2\\
HE 2202$-$4831   &3&56.32&0.67&0.797&  &5331&2.95& $-$2.78 &2.41&8.08& $-$1.28 & $-$2.31 & 2\\
HE 2247$-$7400   &2&5.78&0.59&0.602&  &4829&1.56& $-$2.87 &0.7&6.58& $-$0.94 & $-$0.69 & 2\\
HE 2318$-$1621   &7& $-$41.77 &0.28&0.218&1&4846&1.4& $-$3.67 &1.04&6.3& $-$1.61 & $-$1.34 & 14\\
SDSS J0140$+$2344\tablefootmark{b} &8& $-$198.65 &4.54&$<10^{-6}$&2&5703&4.68& $-$4.00 &1.13&5.56& <$+$0.34 & $-$0.81 & 2\\
SDSS J0929$+$0238   &26&388.33&10.4&&2&5894&4.5& $-$4.97 & 3.91\tablefootmark{a} & 7.44\tablefootmark{a} & <$+$1.46 & $-$3.56\tablefootmark{a} & 15\\
SDSS J1313$-$0019  &3&255.5&13.5&$<10^{-6}$&  2? &5200&2.6& $-$5.00 &2.98&6.41& <$+$0.22 & $-$2.45 & 16\\
SDSS J1341$+$4741   &5& $-$194.68 &26.63&$<10^{-6}$&2&5450&2.5& $-$3.20 & 0.99\tablefootmark{a} & 6.22\tablefootmark{a} & $-$0.73 & $-$0.28\tablefootmark{a} & 17\\
SDSS J1422$+$0031   &6& $-$118.97 &6.09&$<10^{-6}$&2&5200&2.2& $-$3.03 &1.7&7.11& $-$1.18 & $-$0.95 & 18\\
SDSS J1613$+$5309   &4&0.01&0.89&0.745&  &5350&2.1& $-$3.33 &2.09&7.21& $+$0.03 & $-$1.18 & 18\\
SDSS J1746$+$2455   &5&78.45&0.54&0.431&  &5350&2.6& $-$3.17 &1.24&6.51& $+$0.26 & $-$0.56 & 2\\
SDSS J2206$-$0925   &3&14.83&0.9&0.626&  &5100&2.1& $-$3.17 & 0.64\tablefootmark{a} & 5.9\tablefootmark{a} & $-$0.85 & $+$0.61\tablefootmark{a} & 18\\
Segue 1$-$7   &4&204.86&0.59&0.580&  &4960&1.9& $-$3.52 & 2.3\tablefootmark{a} & 7.21\tablefootmark{a} & <$-$0.96 & $+$0.94\tablefootmark{a} & 19\\
SMSS 0313$-$6708   &34&298.5&0.35&0.592&1&5125&2.3& <$-$6.50 & >5.39 &6.02&  & $-$1.20 & 20\\

\hline
\end{tabular}
}

\tablefoot{Columns: star name; number of radial velocity measurements in Table~\ref{table:litrvs}; weighted average of radial velocity measurements in Table~\ref{table:litrvs}; standard deviation of radial velocity measurements in Table~\ref{table:litrvs}; variability criterion $P(\chi^2)$; binary status (1 = single, 2 = binary, labels as collected by \citet{Yoon16} and updated in this work); effective temperature; surface gravity; iron abundance; carbon to iron ratio; absolute carbon abundance; barium to iron ratio, magnesium to carbon ratio; literature reference. \\
\tablefoottext{a}{These carbon abundances have not been corrected for evolutionary status (all other measurements have been, as reported in \citet{Yoon16}). However, all of these stars have $\log g$ $\gtrsim 2.0$, where the correction is expected to be small.} \\
\tablefoottext{b}{Of the two provided solutions in the reference, these are the dwarf solutions.}

\textbf{References:} (1) \citet{Ito13}, (2) \citet{Yong13}, (3) \citet{Jacobson15}, (4) \citet{Giridhar01}, (5) \citet{Roederer14}, (6) \citet{Placco16b}, (7) \citet{PlezCohen05}, (8) \citet{Christlieb04}, (9) H16a, (10) \citet{Norris07}, (11) \citet{Barklem05}, (12) \citet{Frebel08}, (13) \citet{Cohen13}, (14) \citet{Placco14}, (15) \citet{Bonifacio15} and for magnesium \citet{Caffau11}, (16) \citet{Frebel15}, (17) \citet{Bandyopadhyay18}, (18) \citet{Aoki13}, (19) \citet{Norris10}, (20) \citet{Bessell15} and for magnesium \citet{Keller14} \\
}

\end{table*}

\newpage

\begin{table*}[h!]
\caption{\label{table:litrvs} Compilation of 710 radial velocity measurements for 45 CEMP-no stars.}
\begin{tabular}{lrrrll}
 \hline
 name & rv & err & HJD & ref & bibcode \\
   & (km$\,$s$^{-1}$) & (km$\,$s$^{-1}$) & $-2450000$ & & \\
 \hline

BD$+$44$^{\circ}$493 & 	$-150.66$	 & 0.55	& $-4273.25$ &	 \citealt{Carney03}  & 2003AJ....125..293C \\
BD$+$44$^{\circ}$493 &	$-151.16$	 & 0.77	& $-4034.25$ &	 \citealt{Carney03}  & 2003AJ....125..293C \\
BD$+$44$^{\circ}$493 &	$-149.55$	 & 0.68	& $-3698.25$ &	 \citealt{Carney03}  & 2003AJ....125..293C \\
BD$+$44$^{\circ}$493 &	$-151.57$	 & 0.57	& $-3662.25$ & \citealt{Carney03}  & 2003AJ....125..293C \\
BD$+$44$^{\circ}$493 &	$-150.93$	 & 0.67	& $-3634.25$ &	 \citealt{Carney03}  & 2003AJ....125..293C \\
... & ... & ... & ... & ... & ...  \\

\hline
\end{tabular}

\tablefoot{
We do not include \textit{Gaia} DR2 radial velocities in this table because they are averages over multiple measurements in time. \\
(This table is available in its entirety in machine-readable form.) \\
}

\end{table*}

\newpage

\section{Extra figures}

\begin{figure*}[h!]
\centering
\includegraphics[width=0.5\textwidth,trim={3.0cm 0.5cm 2.7cm 0.0cm}]{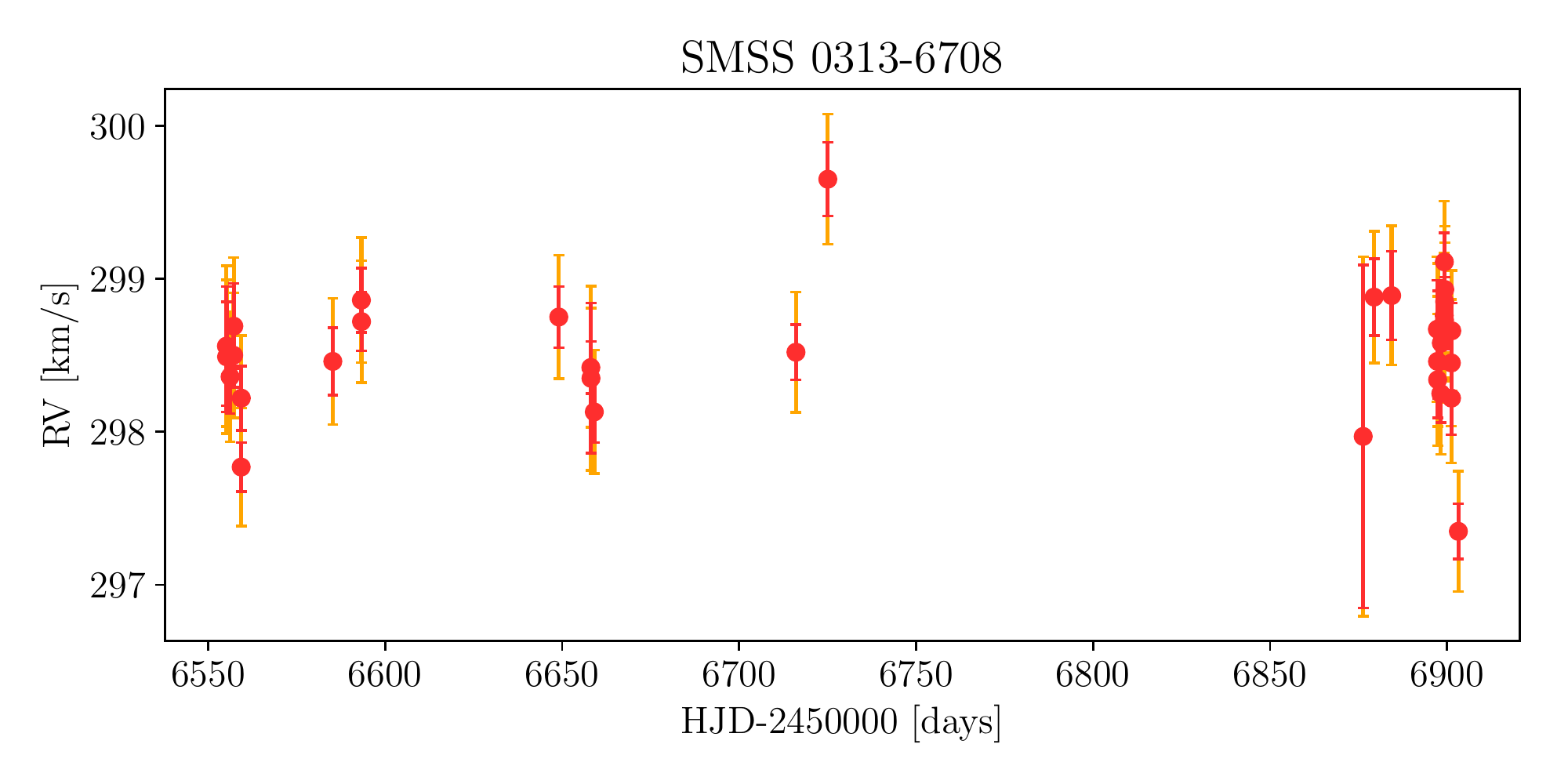}
\caption{Radial velocities for SMSS~0313$-$6708, derived from archive UVES spectra. The red error bars represent the uncertainties coming from \textsc{fxcor}, whereas the orange error bars include an uncertainty floor of 0.35~\kms. }
 \label{fig:SMSS0313}
\end{figure*}

\begin{figure*}[h!]
\centering
\includegraphics[width=\textwidth,trim={3.0cm 1.07cm 2.7cm 0.0cm}]{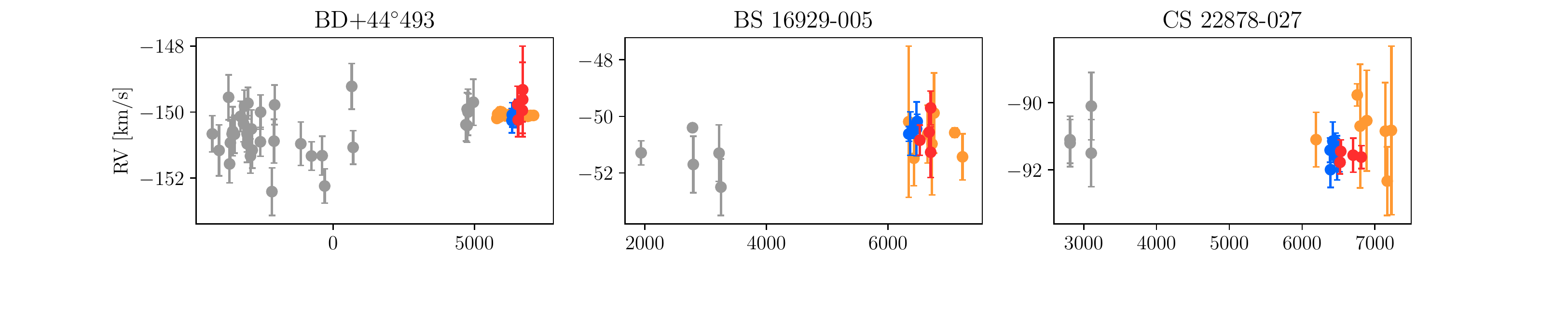}
\includegraphics[width=\textwidth,trim={3.0cm 1.07cm 2.7cm 0.0cm}]{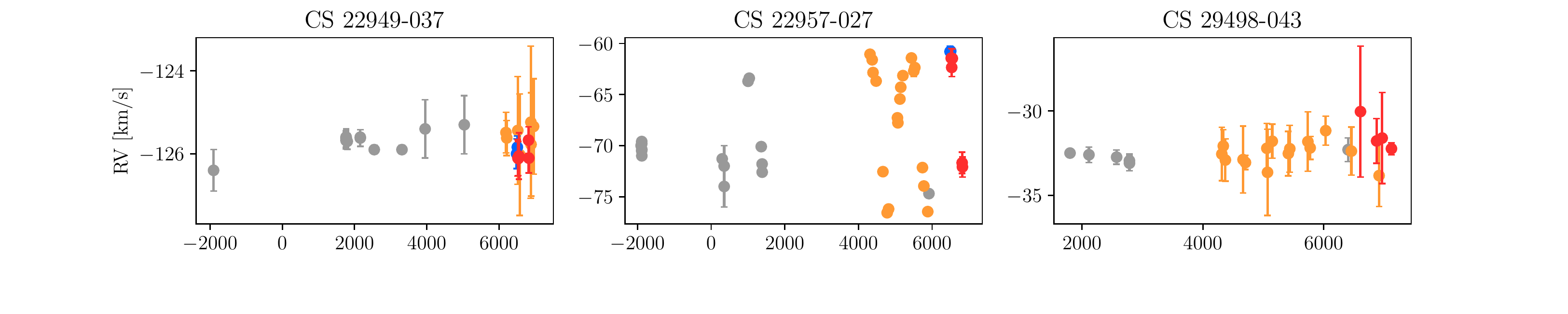}
\includegraphics[width=\textwidth,trim={3.0cm 1.07cm 2.7cm 0.0cm}]{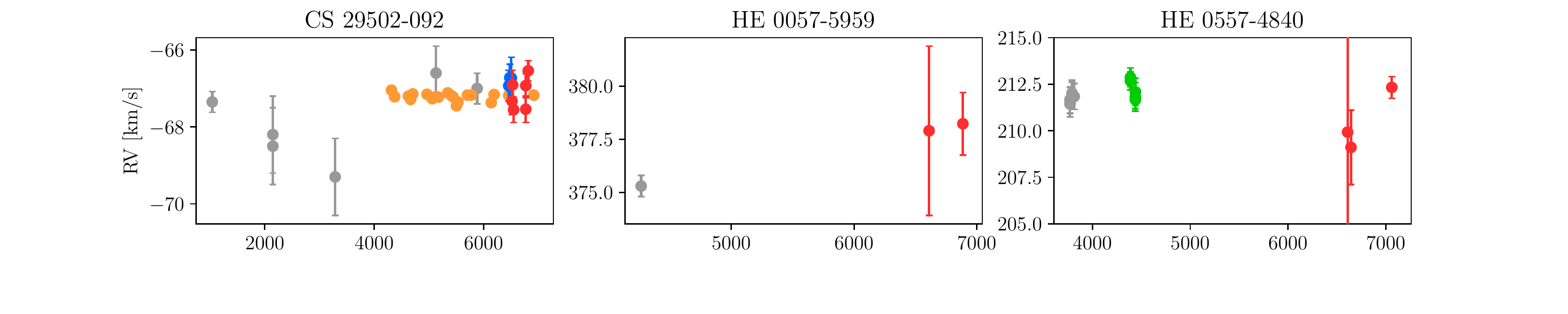}
\includegraphics[width=\textwidth,trim={3.0cm 1.07cm 2.7cm 0.0cm}]{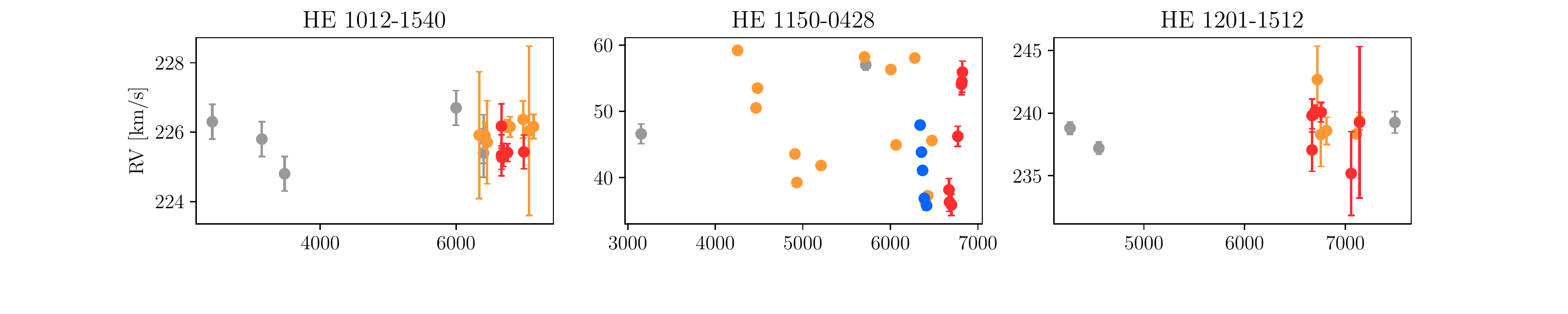}
\includegraphics[width=\textwidth,trim={3.0cm 1.07cm 2.7cm 0.0cm}]{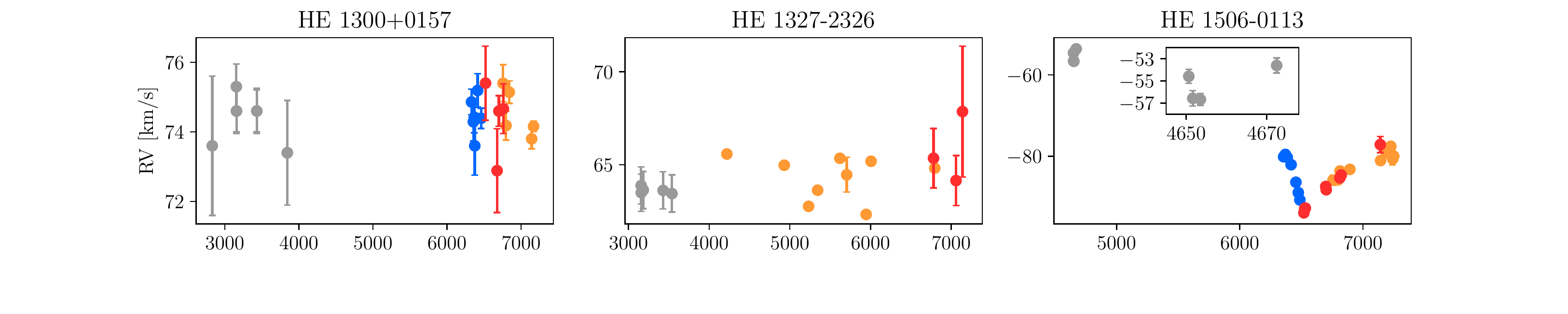}
\includegraphics[width=\textwidth,trim={3.0cm 0.0cm 2.7cm 0.0cm}]{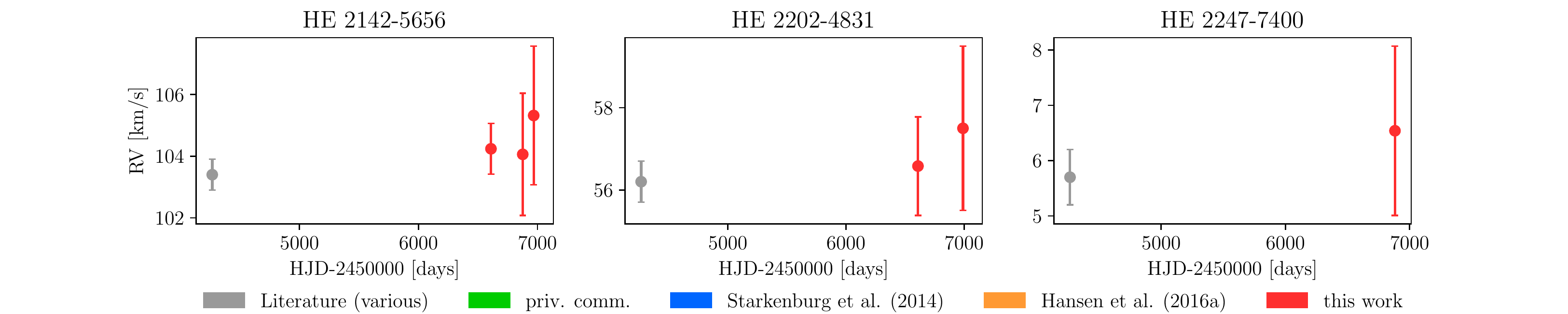}
\caption{Radial velocities of the 18 program stars that have not yet been shown in the main text, ordered alphabetically. For HE~1327$-$2326, there are 7 measurements from \citet{Hansen16a} without provided uncertainties.}
 \label{fig:appendixrvs}
\end{figure*}

\end{appendix}

\end{document}